\documentclass[pra,onecolumn,aps,showpacs,preprint]{revtex4-1}
\usepackage{amsmath}
\usepackage{amssymb,ams fonts,times,natbib}
\usepackage{graphicx}
\usepackage{color}
\usepackage{subfig}

\begin{document}

\title{Nonlinear transmission spectroscopy with dual frequency combs}
\author{Rachel Glenn}
\author{Shaul Mukamel}
\affiliation{Department of Chemistry, University of California, Irvine, California 92697-2025, USA}
\date{\today}
%41.85.Ct	Particle beam shaping, beam splitting
%06.20.-f	Metrology
%42.62.Eh	Metrological applications; optical frequency synthesizers for precision spectroscopy (see also 06.20.-f Metrology in metrology, measurements, and laboratory procedures)
%42.65.Re	Ultrafast processes; optical pulse generation and pulse compression (for ultrafast spectroscopy, see %78.47.J-; for ultrafast magnetization dynamics, see 75.78.Jp)
\begin{abstract}
We show how two frequency combs $\mathcal{E}_1$, $\mathcal{E}_2$ can be used to
measure single-photon, two-photon absorption (TPA),
and Raman resonances in a molecule with three electronic bands,
by detecting the radiofrequency modulation of the nonlinear transmission signal.
Some peaks are shifted by the center frequency of the comb and have a width close to the comb width.
Other peaks are independent of the carrier frequency of the comb.
TPA and Raman resonances that are independent of the carrier frequency
are selected by measuring the transmission signal $\sim\mathcal{E}_1^2 \mathcal{E}_2^2$
and
 the single-photon resonances are selected by measuring the transmission signal
  $\sim\mathcal{E}_1^3\mathcal{E}_2$.
 that interacts three times with comb 1.
 Sinusoidal spectral phase shaping strongly affects  the TPA resonances, but not the Raman resonances,
 for an even or odd phase profile around a selected resonance.
\end{abstract}

\maketitle
%%%%%%%%%%%%%%%%%%%%%%%%%%%%%%%%%%%%%%%%%
%
%
%
%
\section{Introduction}
%
%
%

%%%%%%%%%%%%%%%%%%%%%%%%%%%%%%%%%%%%%%%%%%%

Optical frequency combs, first introduced in 1999\cite{udem_accurate_1999},
have revolutionized meterology\cite{udem_accurate_1999,cundiff_colloquium:_2003}
due to their high resolution of optical frequencies.
They have  been employed for calibrating
sources of spectrographs in astronomy\cite{wilken_spectrograph_2012},
identifying multiple molecules simultaneously\cite{ideguchi_coherent_2013},
 Doppler-free spectroscopy\cite{barmes_high-precision_2013,barmes_spatial_2013,weiner_frequency_2013},
improving energy efficiency in environmental monitoring\cite{giaccari_active_2008},
and forensic analysis, among its many applications. This technology has also enabled the generation of attosecond pulses\cite{krausz_attosecond_2009}.
Because the measurement times of the interferometric signal
can be shortened from seconds using conventional pulse techniques, such as the
scanning-arm Michelson interferometer\cite{femtosecond_book},
  to microseconds with dual-comb;
future possible applications include the observation of chemical reactions in real time\cite{keilmann_mid-infrared_2012}.

Dual comb Fourier transform spectroscopy
\cite{schliesser_mid-infrared_2012,bernhardt_cavity-enhanced_2010,coddington_coherent_2008,coddington_coherent_2010,diddams_direct_2000,hansch_laser_2013,giaccari_active_2008}
 is a technique employed for its
spectral resolution and its concise recording times compared to conventional Fourier transform spectroscopy\cite{hansch_dual_2010}.
It employees two coherent broadband optical frequency combs and
records the nonlinear transmission
in the time domain. A Fourier transformation reveals Raman resonances
in the radio-frequency regime\cite{adler_mid-infrared_2010,ideguchi_adaptive_2014,foltynowicz_quantum-noise-limited_2011, baumann_spectroscopy_2011,bernhardt_mid-infrared_2010, ideguchi_raman-induced_2012,coddington_coherent_2010,bernhardt_cavity-enhanced_2010,hansch_dual_2010}.
Previously, the spectrum was calculated numerically
by means of calculating the intensity of light transmitted through an absorbing gas\cite{foltynowicz_quantum-noise-limited_2011} or by means of fitting with the
nonlinear least-squares method\cite{adler_mid-infrared_2010}.
Here we calculate the third order signal obtained with two frequency combs
and connect them to the third order susceptibility $\chi^{(3)}$. We address several issues:
  how the peaks map from the optical to the radio-frequency regime;
how do the single-photon  and two-photon resonances show up in the
the transmission spectrum;
 how to selectively detect the TPA, Raman, and single-photon resonances in the transmission spectrum;
and can we control these resonances by means of pulse shaping.
We find the positions of some peaks are sensitive
to the carrier frequency of the frequency comb, while other peaks  not sensitive to carrier frequency.
New peaks not studied previously\cite{adler_mid-infrared_2010,ideguchi_adaptive_2014,foltynowicz_quantum-noise-limited_2011, baumann_spectroscopy_2011,bernhardt_mid-infrared_2010, ideguchi_raman-induced_2012,coddington_coherent_2010,bernhardt_cavity-enhanced_2010,hansch_dual_2010}
are calculated.

Pulse-shaping allow the control of the phase $\phi(\omega)$ and amplitude
$\tilde{\mathcal{E}}(\omega)$
of the electric field
$\mathcal{E}(\omega)$
\begin{equation}
\mathcal{E}(\omega)=\tilde{\mathcal{E}}(\omega)e^{i\phi(\omega)}
\end{equation}
and has
inspired the generation of arbitrary waveforms at optical frequencies
\cite{ferdous_spectral_2011,zhou_pair-by-pair_2013,rashidinejad_generation_2013,cundiff_optical_2010}.
We investigate how a sinusoidal phase added to the frequency comb affects the peaks in the spectrum.

This paper is organized as follows. In Sec. II we write the expressions for the nonlinear transmission
spectrum. The transmission signal with of a Lorentzian pulse
is plotted in Sec. III.
The frequency comb in the time and frequency domain
is presented in Sec. IV.
 The selection of the comb line numbers in the transmission spectrum and simulation
 of the transmission spectrum is given in Sections V
 and VI.
 The comb transmission for a sinusoidal spectral phase is simulated in Sec. VII.
The summary is presented in Sec. VIII.

%%%%%%%%%%%%%%%%%%%%%%%%
\section{The Nonlinear transmission signal}
%%%%%%%%%%%%%%%%%%%%%%%%%%

We calculate the transmission signal measured in the time domain
and Fourier transformed
to give the transmission spectrum\cite{mukamelbook}
\begin{equation}
S_t(\omega_s)=-\frac{2}{\hbar}\mathcal{I}\int dt e^{i\omega_s t} \mathcal{E}^*(t)P(t),
\end{equation}
this yields
\begin{equation}
S_{t}(\omega_s)=-\frac{2}{\hbar}
\mathcal{I}
\int d\omega'
\tilde{\mathcal{E}}^*(\omega'-\omega_s) P(\omega'),
%P(\omega)=\int_{-\infty}^{\infty} dt \langle \mathcal{T}V_L (t) e^{-\frac{i}{\hbar}\int_{-\infty}^{\infty}H_{int-}(T)dT}\rangle e^{i\omega t},
\label{freq0}
\end{equation}
where $P(\omega)$ is the polarization induced in the matter by the light and
 $\mathcal{I}A(\omega)$ denotes the imaginary part.
%Equation \eqref{overallS1} represents the signal in the frequency domain.
The polarization will be expanded in powers of the radiation field\cite{mukamelbook}
\begin{eqnarray}
P(\omega)=P^{(1)}(\omega)+P^{(3)}(\omega).
\label{Polarization2}
\end{eqnarray}
The first-order polarization is given by
\begin{equation}
P^{(1)}(\omega)=\mathcal{E}(\omega)\chi^{(1)}(\omega),
\label{P1}
\end{equation}
where $\chi^{(1)}(\omega)$ is the linear susceptibility.
%$\chi^{(1)}(\omega)=\sum_{e_1 g_1}-\frac{1}{\hbar} |\mu_{e_1g_1}|^2 G_{e_1g_1}(\omega)$.
Inserting $P^{(1)}(\omega)$ into  Eq. \eqref{freq0} gives
\begin{equation}
S_{t}^{(1)}(\omega_s)=-\frac{2}{\hbar}
\mathcal{I}
\int d\omega'
\tilde{\mathcal{E}}^*(\omega'-\omega_s) \mathcal{E}(\omega')\chi^{(1)}(\omega').
\label{Si1}
\end{equation}

The third order polarization is given as\cite{mukamelbook}
\begin{eqnarray}
P^{(3)}(\omega)=&&
\int
{d\omega_1}
{d\omega_2}
{d\omega_3}
\mathcal{E}(\omega_1)
\mathcal{E}(\omega_2)
\mathcal{E}^*(\omega_3)
\nonumber\\&&\times
\chi^{(3)}(-\omega;\omega_1,\omega_2,\omega_3)
2 \pi
\delta(\omega-\omega_1-\omega_2+\omega_3),
\label{Polarization4}
\end{eqnarray}
where the susceptibility will depend upon the model of the system.
Inserting Eq. \eqref{Polarization4} into Eq. \eqref{freq0}  gives
\begin{eqnarray}
&&S_{t}^{(3)}(\omega_s )=
-
\frac{2}{\hbar}
\mathcal{I}
\int d\omega'
 d\omega_1
 d\omega_2
 d\omega_3
\tilde{\mathcal{E}}^*(\omega'-\omega_s)
%\int dt
%\tilde{\mathcal{E}}^*(t)
%e^{i(\omega_s-\omega)t}
\tilde{\mathcal{E}}(\omega_2)
\tilde{\mathcal{E}}^*(\omega_3)
\nonumber\\&&\times
\tilde{\mathcal{E}}(\omega_1)
2\pi \delta(\omega'-\omega_1-\omega_2+\omega_3)
\chi^{(3)}(-\omega',\omega_1, \omega_2,\omega_3).
\label{freq00}
\end{eqnarray}
Equation \eqref{freq00} will be used to calculate the transmission
spectrum with a single broadband pulse and with a shaped-pulse composed of two frequency combs.

For comparison we also examine the heterodyne detected signal, i.e., transmission spectrum, measured in the frequency domain,
called the  frequency-dispersed transmission spectrum\cite{mukamelbook}
\begin{equation}
\mathcal{S}_f(\omega)=-\frac{2}{\hbar}
\mathcal{I}
\tilde{\mathcal{E}}^*(\omega) P(\omega).
%P(\omega)=\int_{-\infty}^{\infty} dt \langle \mathcal{T}V_L (t) e^{-\frac{i}{\hbar}\int_{-\infty}^{\infty}H_{int-}(T)dT}\rangle e^{i\omega t},
\label{overallS1}
\end{equation}
Inserting Eq. \eqref{P1}, the first-order signal is given as
\begin{equation}
\mathcal{S}_{f}^{(1)}(\omega)=-\frac{2}{\hbar}
\mathcal{I}
|\tilde{\mathcal{E}}(\omega)|^2 \chi^{(1)}(\omega).
\label{overallS15}
\end{equation}
Unlike Eq. \eqref{Si1}, the signal does not depend upon the phase of the field.

Using  Eq. \eqref{Polarization4}, the
third order dispersed spectrum is given as
\begin{eqnarray}
\mathcal{S}_{f}^{(3)}(\omega)=&&
\mathcal{I}
\frac{2 }{\hbar}
\mathcal{E}^*(\omega)
\int
{d\omega_1}
{d\omega_2}
{d\omega_3}
\mathcal{E}(\omega_1)
\mathcal{E}(\omega_2)
\mathcal{E}^*(\omega_3)
\nonumber\\&&\times
\chi^{(3)}(-\omega;\omega_1,\omega_2,\omega_3)
2 \pi
\delta(\omega-\omega_1-\omega_2+\omega_3).
\nonumber\\
\label{overallS11}
\end{eqnarray}
In the next section, we compare Eqs.  \eqref{freq00} and \eqref{overallS11}
for a single pulse.

%%%%%%%%%%%%%%%%%%%%%%%%%%%%%%%%%%%%%%%%%%%%%%
%
%
\section{ transmission signal of a Broadband pulse}
%
%
%
%%%%%%%%%%%%%%%%%%%%%%%%%%%%%%%%%%%%%%%%%

%%%%%%%%%%%%%%%%%%%%%%%%%%%%%%%%%%%%%%%%
\begin{figure}[t]
\begin{center}
\includegraphics[scale=0.35]{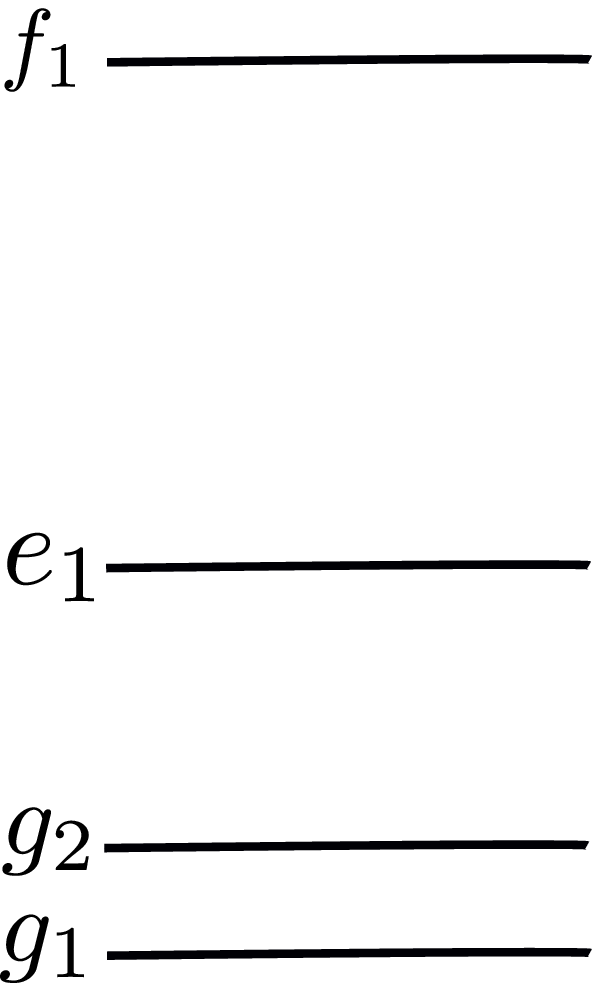}
\end{center}
\caption{The model level scheme contains three electronic states with the
transition frequencies
$\omega_{f_1 g_1}=36,000 \mathrm{cm}^{-1}$,
$\omega_{e_1 g_1}=12,000 \mathrm{cm}^{-1}$,
$\omega_{g_2 g_1}=1,200 \mathrm{cm}^{-1}$.
The dephasing rates are
$\Gamma_{f_1 g_1}=500\mathrm{cm}^{-1}$,
$\Gamma_{e_1 g_1}=100 \mathrm{cm}^{-1}$,
$\Gamma_{g_2 g_1}=80 \mathrm{cm}^{-1}$.
The transition dipole moments are set to one.
}
\label{fig:level_scheme}
\end{figure}
%%%%%%%%%%%%%%%%%%%%%%%%%%%%%%%%%%%%%%%%%%%%%%%%%%%

%%%%%%%%%%%%%%%%%%%%%%%%%%%%%%%%%
\begin{figure*}[t]
\begin{center}
\includegraphics[scale=0.55]{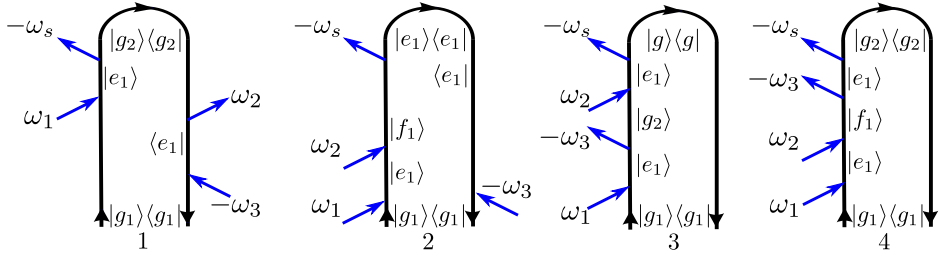}
\end{center}
\caption{(Color online) Loop diagrams for the transmitted signal
Eq.  \eqref{freq00} or Eq. \eqref{overallS11}.
}
 \label{fig:overall}
\end{figure*}
%%%%%%%%%%%%%%%%%%%%%%%%%%%%%%%%%%%%%%%%%

%%%%%%%%%%%%%%%%%%%%%%%%%%%%%%
We consider a three band model system  Fig. \ref{fig:level_scheme}  with electronic states
$|g\rangle$,
$|e\rangle$,
$|f\rangle$.
% Here $|g\rangle$ is the ground state.
The linear susceptibility then reads\cite{mukamelbook}
\begin{equation}
\chi^{(1)}(\omega)=\sum_{e_1 g_1}-\frac{1}{\hbar} |\mu_{e_1g_1}|^2 G_{e_1g_1}(\omega),
\end{equation}
where $G_{e_1 g_1}(\omega)=(\omega-\omega_{e_1 g_1}+i\Gamma_{e_1g_1})^{-1}$.

 The third-order susceptibility can be read off the diagrams of Fig. \ref{fig:overall} \cite{mukamelbook}
\begin{eqnarray}
&&
\chi^{(3)}
(-\omega;\omega_1,\omega_2,\omega_3)=
\left(\frac{-1}{2\pi \hbar}\right)^3
%\frac{1}{3!}\sum_p
\sum_{g_i, e_i,f_i}
V_{g_1 e_1}
V_{ e_1 g_2}
V_{  g_2 e_1}
V_{ e_1 g_1}
\nonumber\\&&\times
G_{e_1}^*(-\omega+\omega_1+\omega_2)
G_{g_2}^*(-\omega+\omega_1)
G_{e_1}(\omega_1)
+
V_{g_1 e_1}
V_{ e_1 f_1}
\nonumber\\&&\times
V_{  f_1 e_1}
V_{ e_1 g_1}
G_{e_1}^*(-\omega+\omega_1+\omega_2)
G_{f_1}(\omega_1+\omega_2)
G_{e_1}(\omega_1)
\nonumber\\&&+
V_{g_1 e_1}
V_{ e_1 g_2}
V_{  g_2 e_1}
V_{ e_1 g_1}
G_{g_2}(\omega_1-\omega_3)
G_{e_1}(\omega_1-\omega_3+\omega_2)
\nonumber\\&&\times
G_{e_1}(\omega_1)
+
V_{g_1 e_1}
V_{ e_1 f_1}
V_{  f_1 e_1}
V_{ e_1 g_1}
G_{e_1}(\omega_1+\omega_2-\omega_3)
\nonumber\\&&\times
G_{f_1}(\omega_1+\omega_2)
G_{e_1}(\omega_1).
\label{S4}
\end{eqnarray}

\begin{figure}[t]
\begin{center}
\includegraphics[scale=0.25]{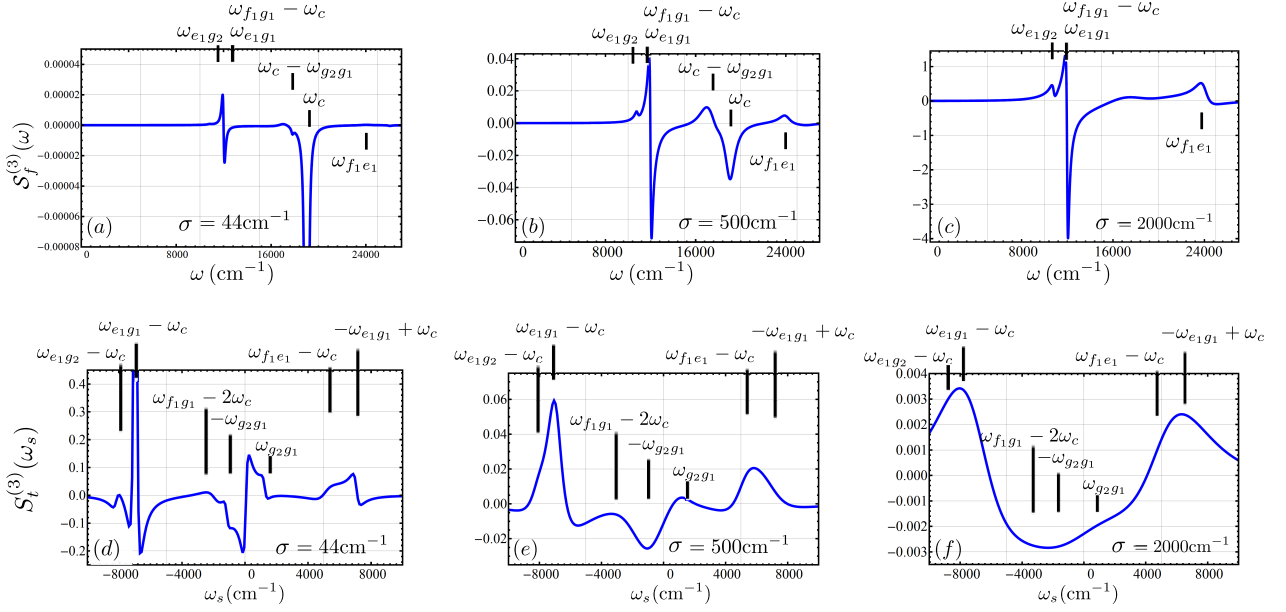}
\end{center}
\caption{(Color online)
Top: The frequency-dispersed transmission spectrum
  $\mathcal{S}_f^{(3)}(\omega)$  Eq. \eqref{overallS11} is plotted for a
Lorentzian pulse Eq. \eqref{Gaussian1},
for several values of $\sigma$.
Bottom: The Fourier transform of the time-resolved transmission signal $S_t^{(3)}(\omega_s)$ Eq. \eqref{freq00}
is plotted  for a Lorentzian pulse Eq. \eqref{Gaussian1},
for several values of $\sigma$.
The signal is plotted in the units $ \frac{2 \pi}{\hbar}(\frac{1}{2\pi10,000 \hbar})^3$
and the carrier frequency is  $\omega_c=19000  \mathrm{cm}^{-1}$.
All dipole moments are set to one.
%To evaluate
%the summation we randomly sample the two summations over $n$ and $m$. Here
%we sample 2000, or 3500 different pulses in the range $n$, $m$$<22250\mathrm{cm}^{-1}/\omega_{rep}=370,833$.
}
 \label{fig:Plot0}
\end{figure}

The frequency-dispersed transmission spectrum Eq. \eqref{overallS11}
with a Lorentzian pulse\cite{dorfman_multidimensional_2014}
\begin{equation}
\mathcal{E}(\omega)=
\frac{\sigma}{\omega+i\sigma}
%e^{-\omega^2 /2\sigma^2}
\label{Gaussian1}
\end{equation}
is calculated analytically and shown in the top row of Fig. \ref{fig:Plot0}.
%Note that we use a Lorentzian pulse in this section. In the following sections,
%we will use a Gaussian shaped pulse.
 $\mathcal{S}_f^{(3)}(\omega_s)$ is plotted in the units $ \frac{2 \pi}{\hbar}(\frac{1}{2\pi10,000 \hbar})^3$
 with the dipole moments set to one.
 In Fig. \ref{fig:Plot0}(a), the resonances  $\omega=\omega_{f_1 e_1}$, $\omega_{e_1 g_1}$,
 $\omega_{e_1 g_2}$ are marked. The transmission spectrum contains other peaks
 such as $\omega_s=\omega_c-\omega_{g_2 g_1}$,
 %$\omega_{e_1 g_1}+2\omega_c$,
 $\omega_{f_1 g_1}-\omega_c$. The   $\omega_s=\omega_{f_1 g_1}-\omega_c$ peak
 overlaps with the $\omega_s=\omega_{e_1 g_1}$ peak.
 %The $\omega_s=\omega_c-\omega_{g_2 g_1}$
 %peak is small can be seen in the inset.
 % The $\omega_s=\omega_{e_1 g_1}+2\omega_c$ peak lies out
 %outside the range.
 The dominate peak in the transmission spectra is the
peak at the carrier frequency $\omega_s=\omega_c$. As the pulse width increases,
in Fig.  \ref{fig:Plot0}(b), the $\omega_s=\omega_{f_1 g_1}$, $\omega_{e_1 g_2}$ peaks become seen and the $\omega_c$ peak decreases. Increasing the
pulse width further, Fig.  \ref{fig:Plot0}(c), these peaks become more pronounced.

 The Fourier transform of the time-resolved transmission signal Eq. \eqref{freq00} is shown the bottom row of Fig. \ref{fig:Plot0}
 for the electric field \eqref{Gaussian1}.
In Fig.\ref{fig:Plot0}(a), the two-photon transition $\omega_{f_1 g_1}$ interacts two times with the pulse and it
is shifted by 2$\omega_c$. The single-photon transitions,  $\omega_{e_1 g_1}$, $\omega_{f_1 e_1}$,
 interact once with pulse and are shifted by $\omega_c$.
The Raman peaks  are not shifted, since they interact twice with the pulse, once
with $\omega_c$ and a second with $-\omega_c$, which cancels.
%There is a peak at $\omega_s=0$, which is a result of  four interactions with
%the same pulse.
Increasing the pulse width to $\sigma=500\mathrm{cm}^{-1}$ in  Fig.  \ref{fig:Plot0}(e) the $\omega_s=\omega_{e_1g_1}-\omega_c$
peak remains dominate. This is also true for Fig. \ref{fig:Plot0}(b). Increasing the pulse width further,
in  Fig.  \ref{fig:Plot0}(f) The peaks become smeared. Overall, the
two signals $\mathcal{S}_{f}^{(3)}(\omega)$ and $S_t^{(3)}(\omega_s)$
  are different.

%%%%%%%%%%%%%%%%%%%%%%%%%%%%%%%%%%

%%%%%%%%%%%%%%%%%%%%%%%%%%%%%%%%%%%%%%%%%%%%%%%%%%%%%%
%
%
%
%
\section{The Dual Frequency comb}
%
%
%
%
%
%%%%%%%%%%%%%%%%%%%%%%%%%%%%%%%%%%%%%%%%%%%%%%%%%%%%%%%%%%%%

The frequency comb is  generated using a mode-locked laser that produces a series of optical pulses separated by the round-trip
time of the laser cavity $T_{rep}=l_c /v_g$, where $v_g$ is the group velocity and $l_c$ is the round-trip
length of the laser cavity\cite{schliesser_mid-infrared_2012,rauschenberger_control_2002,holman_detailed_2003,jones_carrier-envelope_2000}
%\cite{ideguchi_raman-induced_2012}
.
We consider two femtosecond frequency combs, which with the electric field $E(t)=\mathcal{E}(t)+\mathcal{E}^*(t)$
\begin{eqnarray}
&&\mathcal{E}(t)=
e^{-i \omega_{c} t}
\sum_{n=1}^{N}
\tilde{\mathcal{E}}(t+n T_{rep,1}) e^{i n (\omega_{ceo,1}+\omega_c) T_{rep,1}}
\nonumber\\&&+
e^{-i \omega_{c} (t-\Delta t)}
\sum_{n=1}^{N}
\tilde{\mathcal{E}}(t+n T_{rep,2}-\Delta t) e^{i n (\omega_{ceo,2}+\omega_c) T_{rep,2}},
\label{intro1}
\nonumber\\&&
\end{eqnarray}
%separated by a repetition time, $T_{rep}$,
where $\omega_c$ is the carrier frequency and $\Delta t$ is the delay between the two combs.
The summation index $n$ represents the pulse number with a total of $N$ pulses.
The envelope function $\tilde{\mathcal{E}}(t)$ is periodic $\tilde{\mathcal{E}}(t)=\tilde{\mathcal{E}}(t+n T_{rep})$.
The repetition frequencies $\omega_{rep,1}=2\pi /T_{rep,1}$ and
$\omega_{rep,2}=2\pi /T_{rep,2}$ are close, such that
 $\delta\omega_{rep}\ll \omega_{rep,1}$,  where
$\delta \omega_{rep}=\omega_{rep,1}-\omega_{rep,2}$.

$\omega_{ceo,i}= (\Delta \phi /2\pi) \omega_{rep,i}$ is the carrier-envelope offset frequency.   $\Delta \phi=(1/v_g-1/v_p) l_c \omega_c$, is the phase shift between the peak of the
envelope and the closest peak of the carrier wave and   $v_p$
is the phase velocity.
The range of the carrier-envelope phase is $0<\Delta \phi<2\pi$.
It is possible to lock $\omega_{ceo.i}$ to zero\cite{holman_detailed_2003}. We assume a
vanishing carrier-offset frequency $\omega_{ceo,1}= \omega_{ceo,2}=0$.

The frequency comb can be generated by replacing the cavity with with a Fabry-P{\'e}rot etalon, \cite{hoffman_optimally_2013}.
In this method the individual pulse-shape in the pulse train becomes asymmetric.
%which would be advantageous for studying short-time dynamics.
An intracavity etalon is typically employed  for self-stabilization of the optical frequencies and the
pulse repetition rate in conventional frequency comb generation
 with high repetition rates 10 GHz\cite{quinlan_self-stabilization_2008}.
An external molecular absorption cell can also be employed to stabilize the optical frequencies and the optical repetition rate\cite{nakazawa_c2h2_2008}.

An ideal frequency comb uses an infinite train of pulses ($N\rightarrow \infty$) and
the electric field can be represented as a Fourier series,
\begin{eqnarray}
\mathcal{E}(t)
&&=\mathcal{E}_1(t)+\mathcal{E}_2(t-\Delta t)
\nonumber\\&&
=e^{-i \omega_{c}t}
\sum_{n=-\infty}^{\infty} A_{n,1} e^{-i
%\omega_{ceo,1}
n \omega_{rep, 1}t}
%\nonumber\\&&
%\tilde{ \mathcal{E}}_2(t-\Delta t)=
\nonumber\\&&+
e^{-i \omega_{c}(t-\Delta t)}
\sum_{m=-\infty}^{\infty} A_{m,2} e^{-i
%\omega_{ceo,2}
m \omega_{rep,2}(t-\Delta t)},
\label{beating}
\end{eqnarray}
where $A_{n,i}$
\begin{equation}
A_{n,i}=\frac{1}{T_{rep,i}} \int_{-\infty}^{\infty} \mathcal{E}_i(t)e^{- i (n \omega_{rep,i} -\omega_c)t} dt,
\end{equation}
 $\mathcal{E}_i(t)$ is the pulse envelope, the index $i$ represents comb 1 or comb 2.
% For
%$N\rightarrow \infty$, the form of Eq. \eqref{intro1} is identical to Eq. \eqref{beating}.

\begin{figure*}[t]
\begin{center}
\includegraphics[scale=0.45]{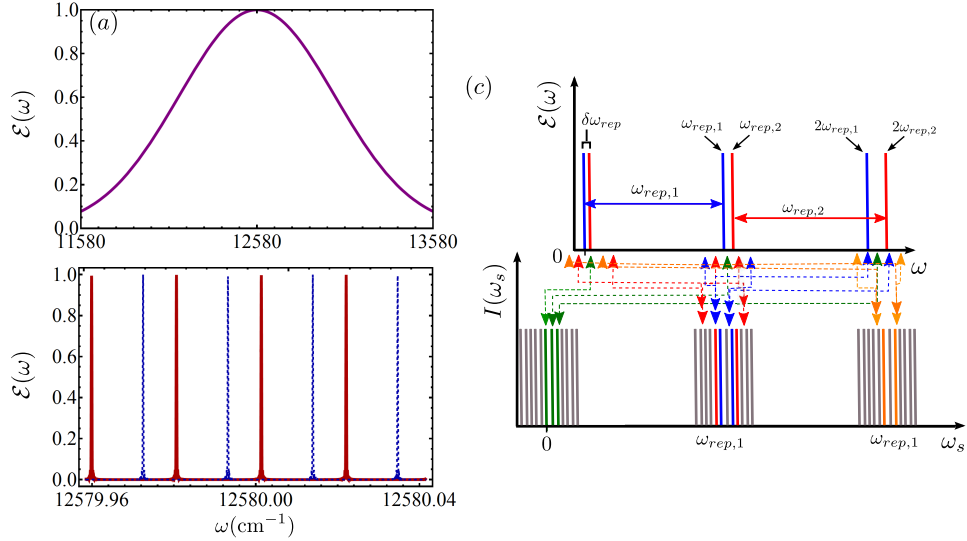}
\end{center}
\caption{(Color online)(a)
Frequency comb, Eq. \eqref{freq38},
for a Gaussian envelope Eq. \eqref{Gaussian}
$\omega_{rep,1}=0.0033\mathrm{cm}^{-1}$,
$\eta=10^{-6}$,
$\omega_c=12580\mathrm{cm}^{-1}$,
and
$\sigma=441\mathrm{cm}^{-1}$.
%There are 315,416 comb lines within the marked FWHM.
(b) The dual comb in (a) is displayed on a much smaller scale
to shown the individual comb lines of comb 1 (blue) and comb 2 (red).
(c) Schematic of $\mathcal{E}(\omega)$ for the dual freq comb Eq. \eqref{freq38}.
(d) Interferogram of the double comb Eq. \eqref{fourieromega} is shown for the first three frequency combs
resulting from the beating of two frequency combs in the time domain.
}
\label{fig:comb}
\end{figure*}

The Fourier transform
$ \mathcal{E}(\omega)=\int_{-\infty}^{\infty}\tilde{ \mathcal{E}}(t) e^{i\omega t}dt$
 of Eq. \eqref{intro1}  produces a frequency  comb
\begin{eqnarray}
\mathcal{E}(\omega)=&&
\tilde{\mathcal{E}}(\omega-\omega_c)
\sum_n^N
e^{in \omega T_{rep,1}+ in \Delta \phi}
\nonumber\\&&+
\tilde{\mathcal{E}}(\omega-\omega_c)
\sum_m^N
e^{im \omega T_{rep,2}+ im \Delta \phi-i\omega\Delta t}
\label{intro2}
\end{eqnarray}
%which a series of comb lines with separation $\omega_{rep}=2\pi / T_{rep}$.
with comb envelope $\tilde{\mathcal{E}}(\omega)=\int_{-\infty}^{\infty} \tilde{\mathcal{E}}(t)e^{-i\omega t} dt$. The summation of the exponentials in Eq. \eqref{intro2} is Fourier series
%that creates a frequency comb structure.
with constructive interference occurring at $\omega T_{rep,i}-\Delta \phi=2 \pi n$.
The center frequency of line number $n$, with
$\omega\rightarrow\omega_n$ is expressed as   $\omega_n=  n \omega_{rep,i}+\omega_{ceo,i}$.
As the number of pulses $N$ is increased the spectral width of the comb lines narrows and for
$N\rightarrow \infty$ Eq. \eqref{intro2} can be simplified as
\begin{eqnarray}
&&\mathcal{E}(\omega)=
\omega_{rep,1}
\tilde{ \mathcal{E}}_1(\omega-\omega_c)
\sum_{n=-\infty}^{\infty}
\delta (n\omega_{rep,1}-\omega)
\nonumber\\&&+
\omega_{rep,2}
\tilde{ \mathcal{E}}_2(\omega-\omega_c)
e^{-i\omega \Delta t}
\sum_{m=-\infty}^{\infty}
\delta (m\omega_{rep,2}-\omega).
\label{freq38}
%\label{freq3}
\end{eqnarray}
Equation \eqref{freq38} is plotted in Fig. \ref{fig:comb} for a
Gaussian envelope
\begin{eqnarray}
&&\tilde{\mathcal{E}}_1(\omega-\omega_c)=E_1 e^{-(\omega-\omega_c)^2/2\sigma^2},
\nonumber\\&&
\tilde{\mathcal{E}}_2(\omega-\omega_c)=E_2 e^{-(\omega-\omega_c)^2/2\sigma^2},
\label{Gaussian}
\end{eqnarray}
with $\sigma=441\mathrm{cm}^{-1}$, $\omega_c=12580\mathrm{cm}^{-1}$ and  the delta-function
is
replaced by
a Lorentzian function.
%$\sigma_\omega=\sqrt{\ln(2)}/(\pi \sigma_t)*3.3356*10^{-11}\mathrm{cm}^{-1}$
Fig. \ref{fig:comb}(a) shows the Gaussian envelope of the two overlapping frequency combs.
There are $315,416$ pulses contained in the full width half max (FWHM).
Fig. \ref{fig:comb}(b) shows the equidistant comb lines of the two frequency combs, in blue and red,
for
$\omega_{rep,1}=0.033\mathrm{cm}^{-1}$ and $\delta\omega_{rep}=10^{-6}\omega_{rep,1}$.

%%%%%%%%%%%%%%%%%%%%%%%%%%%%%%%%%%%%%%%%
%%%%%%%%%%%%%%%%%%%%%%%%%%%%%%%%%%%%%%%%%%%%%%%%%%%%%%%%%%%%%%%%%%%%%%%%%%%%%%%%%

The beating of the two combs Eq. \eqref{beating} creates a
time resolved  interferometric signal
%There are an infinite number of beating modes between combs.
$I(t)= | \mathcal{E}(t)|^2$, which reads
\begin{eqnarray}
I(t)=&&
 \big| \mathcal{E}(t)|^2
=
\sum_{p,r} A_{p,1} A^*_{r,1}e^{i (p-r) \omega_{rep, 1}t}
\nonumber\\&&+
\sum_{p,r} A_{p,2} A^*_{r,2}e^{i(p-r) \omega_{rep,2}(t+\Delta t)}
\nonumber\\&&
+
e^{-i \omega_{c}\Delta t}
\sum_{n,m} A_{n,1} A^*_{m,2}
e^{i (n \omega_{rep,1}-m \omega_{rep,2} )t+ im\omega_{rep,2}\Delta t}
\nonumber\\&&
+
e^{i \omega_{c}\Delta t}
\sum_{n,m} A^*_{n,1} A_{m,2}
e^{-i (n \omega_{rep,1}-m \omega_{rep,2})t-im \omega_{rep,2}\Delta t}.
\label{beat}
\nonumber\\&&
\end{eqnarray}
The last two terms in Eq. \eqref{beat} contain
many possible beat frequencies: $n \omega_{rep,1}-m \omega_{rep,2}$.
%In the frequency
%domain, these frequencies combine and give a new frequency.
The Fourier transform of Eq. \eqref{beat}  reads
\begin{equation}
I(\omega_s)=\int dt I(t) e^{i\omega_s t}.
\label{fourieromega}
\end{equation}
For $n=m$, Eq. \eqref{fourieromega}
 will give a frequency comb
$\sum_n\delta(\omega-n\delta\omega_{rep})$.
The application of a second comb, thus down-converts comb 1 by the
factor
\begin{equation}
\eta=
\delta \omega_{rep}/\omega_{rep,1}.
\end{equation}
This frequency comb has line spacing $\delta\omega_{rep}$ and has envelope  is the product of the envelopes of the two fields.

The dual  frequency comb, Eq. \eqref{freq38} is sketched in
Fig. \eqref{fig:comb}(c). Fig. \eqref{fig:comb}(d) sketches the
Fourier transform of the interferometric signal Eq. \eqref{beat} given by Eq. \eqref{fourieromega}.
The first group of lines corresponds to the selection of the modes $n=m$.
The second group corresponds to $m=n+1$ and has the form
$\sum_n\delta(\omega_s-\omega_{rep,2}-n\delta\omega_{rep})$.
It is centered at $\omega_s\approx\omega_{rep,1}$
with line spacing $\delta\omega_{rep}$ and is identical to the first group.
The third group is at  $\omega_s\approx2\omega_{rep,1}$ and corresponds
to the combination $m=n+2$.
The spectrum contains an infinite number of identical frequency combs center at
$\omega_s\approx p \omega_{rep,1}$, where $p$ is an integer.
%All spectral information is contained
%in the first frequency comb for $n=m$.
Typically the only the first group of lines is measured and
the higher frequencies can be cut-off experimentally by
using a low-pass filter in
the acquisition circuit\cite{ideguchi_coherent_2013}.
For two combs with THz carrier frequencies, and repetition frequencies $\omega_{rep,1}= 2 \pi 100$ MHz,
$\delta \omega_{rep}= 2 \pi 100$ Hz, the peaks in the spectrum
are multiplied by $\eta=10^{-6}$ and the spectrum
lies in the radiofrequency
%is down-converted by the factor $\delta\omega_{rep}/\omega_{rep,1}=10^{-6}$ into the radiofrequency
regime\cite{ideguchi_coherent_2013}.
For unambiguous assignment of the comb modes, the bandwidth should not exceed $\pm\omega_{rep,1}/2$,
which can be derived from the Nyquist theorem.

%%%%%%%%%%%%%%%%%%%%%%%%%%%%%%%%%%%%%%%%%%%
%
%
%
\section{Comb line selection in the nonlinear time-resolved transmission signal with scaling $\tilde{\mathcal{E}}_1^2\tilde{\mathcal{E}}_2^2$}
\label{two}
%
%
%%%%%%%%%%%%%%%%%%%%%%%%%%%%%%%%%%%%%%%%%%%

The time-resolved transmission spectrum for two frequency combs contains
the signal $\tilde{\mathcal{E}}_1^2\tilde{\mathcal{E}}_2^2$,
 $\tilde{\mathcal{E}}_1^3\tilde{\mathcal{E}}_2$ and $\tilde{\mathcal{E}}_1\tilde{\mathcal{E}}_2^3$.
We analyze the spectrum separately for
$\tilde{\mathcal{E}}_1^2\tilde{\mathcal{E}}_2^2$ and $\tilde{\mathcal{E}}_1^3\tilde{\mathcal{E}}_2$.
The expression for the spectrum scaling as $\tilde{\mathcal{E}}_1\tilde{\mathcal{E}}_2^3$  are similar to
$\tilde{\mathcal{E}}_1^3\tilde{\mathcal{E}}_2$ with the $\delta \omega_{rep}\rightarrow -\delta \omega_{rep}$.

We select terms that scale as $\tilde{\mathcal{E}}_1^2\tilde{\mathcal{E}}_2^2$.
%involving two interactions with each comb, this can be done by using different amplitudes for the pulses of comb 1 and 2.
Fourier transform of interferometric signal with two interactions from comb 1 and 2, gives the following possible beat frequencies
\begin{eqnarray}
\omega_s&&=(n-r)\omega_{rep,1}+(m-p)\omega_{rep,2},
\nonumber\\&&
=(n+r)\omega_{rep,1}-(m+p)\omega_{rep,2},
\nonumber\\&&
=(n-r)\omega_{rep,1}-(m-p)\omega_{rep,2}.
\label{rep}
\end{eqnarray}
Note that the exchange of $\omega_{rep,1}$ and $\omega_{rep,2}$ is  possible in Eq. \eqref{rep}.
The two interactions with comb 1 correspond the indices $r$ and $n$, and two interactions with comb 2 to
 $p$ and $m$.
%The repetition frequency of comb 1 and comb 2 are close,
% $\delta \omega_{rep}=\omega_{rep,2}-\omega_{rep,1}$.
%with $\delta \omega_{rep} \ll \omega_{rep,1}$.
Similar to the interferometric signal Eq. \eqref{beat},
 the relation $m-p=r-n$, for the first term in Eq. \eqref{rep},
  will give a frequency comb
$\sum_{n,r}\delta((n-r)\delta\omega_{rep}-\omega_s)$.
% $\delta \omega_{rep}/\omega_{rep,1}$.
%and centered at $\delta \omega_{rep}$.
The combination $m-p=r-n+1$ will give identical frequency comb
$\sum_{n,r}\delta((n-r)\delta\omega_{rep}+\omega_{rep,2}-\omega_s)$,
 centered
at $\omega_s\approx\omega_{rep,1}$.
Based on this observation, we use a delta-function
to select the correct combination of line numbers. For example,
the combination of the line numbers in Eq. \eqref{rep}, will acquire the corresponding delta-functions
\begin{eqnarray}
&&
\delta(n-r+m-p),
\nonumber\\&&
\delta(n+r-m-p),
\nonumber\\&&
\delta(n-r-m+p),
\label{combo}
\end{eqnarray}
respectively.
When expanding the field correlation functions we can insert the corresponding delta-function
and eliminate one of the summations over the spectral line numbers.
This is done in Appendix A
and the final expression for the time-resolved transmission spectrum is given in Eq. \eqref{Scombo}.

The time-resolved transmission spectrum $S_t(\omega_s)$, Eq. \eqref{Scombo} contains many peaks. The  TPA and
Raman peaks that do not depend on $n$ or $m$:
 \begin{eqnarray}
&&
\omega_s=\pm\eta\omega_{f_1g_1},
\nonumber\\&&
\omega_s=\pm\eta\omega_{g_2g_1}.
\label{TPARaman}
\end{eqnarray}
Other peaks that depend upon $n$ and $m$
and that lie within the displayed regime $\pm\omega_{rep,1}/2$ are
 \begin{eqnarray}
&&
\tilde{\omega}_{f_1g_1}^+=\eta (\omega_{f_1g_1}-(m+n)\omega_{rep,1})
%\nonumber\\&&
%\tilde{\omega}_{f_1g_1}^-=-\eta (\omega_{f_1g_1}+(m+n)\omega_{rep})
\nonumber\\&&
\tilde{\omega}_{e_1g_1}^+=\eta(\omega_{e_1g_1}-n \omega_{rep,1})
\nonumber\\&&
\tilde{\omega}_{e_1g_1}^-=-\eta(\omega_{e_1g_1}+n \omega_{rep,1})
\nonumber\\&&
\tilde{\omega}_{g_2g_1}^+=\eta (\omega_{g_2g_1}-(m+n)\omega_{rep,1})
\nonumber\\&&
\tilde{\omega}_{g_2g_1}^-=-\eta (\omega_{g_2g_1}+(m+n)\omega_{rep,1}).
\label{peaksfig}
 \end{eqnarray}
The peaks   $\omega_s=-\eta(\omega_{f_1g_1}+(m+n)\omega_{rep,1})$,
$\omega_{e_1g_1}\pm n \omega_{rep,1}$, $\omega_{g_2g_1}+(m+n)\omega_{rep,1}$
lie outside the displayed regime.
The center position of the peaks that depend on $n$ and $m$ can be found by substituting $n=m=\omega_c/\omega_{rep,1}$.
The single-photon peaks $\tilde{\omega}_{e_1g_1}^-$ and $\tilde{\omega}_{e_1g_1}^+$
depend $n$; while, the range of $n$ depends upon the width of the frequency comb.
Hence, the width of these peaks will be close to the width of the frequency comb multiplied by $\eta$.
The TPA and Raman resonances depend on $m$ and $n$, so that these peaks
will be twice as broad as the single-photon peaks.

The  down shifting the peaks can be understood by comparing Eqs. \eqref{freq00}
and  \eqref{overallS11} . Eq. \eqref{freq00} contains an  $\omega'$ integration,
 which is a result of the time resolved signal detection. This integration  mixes the frequency combs and shifts the
peaks into the radiofrequency range. Using the delta function in Eq. \eqref{freq00},
and inserting it  into the field $\tilde{\mathcal{E}}^*(\omega'-\omega_s)$, we  find
$\mathcal{E}^*(\omega_1+\omega_2-\omega_3-\omega_s)$, which mixes the four frequencies in the
diagrams of Fig. \ref{fig:overall}. In Eq. \eqref{overallS11}, using the delta function we have
$\mathcal{E}^*(\omega_1+\omega_2-\omega_3)$, which mixes three of the frequencies.

The modulation of the transmission signal in the radiofrequency range,
can be seen from the expression
Eq. \eqref{Scombo}, which is proportional to
\begin{widetext}
\begin{eqnarray}
&&S_{t1122}^{(3)}(\omega_s; \omega_{rep,1}, \omega_{rep,2},\tau_2)
\propto
-\mathcal{I}
\frac{2}{2\pi\hbar}
\omega_{rep,1}^2
\omega_{rep,2}^2
\Big[
\tilde{\mathcal{E}}_2^*(m\omega_{rep,2}-\omega_c)
\tilde{\mathcal{E}}_1(n\omega_{rep,1}-\omega_c)
\tilde{\mathcal{E}}_2^*(p\omega_{rep,2}-\omega_c)
%((r+n)\omega_{rep,1}-p\omega_{rep,2}-\omega_s-\omega_c)
\nonumber \\&&\times
\tilde{\mathcal{E}}_1(r\omega_{rep,1}-\omega_c)
V_{g_1 e_1}
V_{ e_1 f_1}
V_{  f_1 e_1}
V_{ e_1 g_1}
\nonumber\\&&\times
\frac{\delta((r+n)\omega_{rep,1}-(p+m)\omega_{rep,2}-\omega_s)}
%\delta(n+r-m-p)}
{(r\omega_{rep,1}-\omega_{e_1g_1}+i\Gamma_{e_1})
((n+r)\omega_{rep,1}-\omega_{f_1g_1}+i\Gamma_{f_1})
((n+r)\omega_{rep,1}-p\omega_{rep,2}-\omega_{e_1g_1}-i\Gamma_{e_1})}
\nonumber\\&&+
\tilde{\mathcal{E}}_1^*(n\omega_{rep,1}-\omega_c)
%(r\omega_{rep,1}-(p-m)\omega_{rep,2}-\omega_s-\omega_c)
\tilde{\mathcal{E}}_2(m\omega_{rep,2}-\omega_c)
\tilde{\mathcal{E}}_2^*(p\omega_{rep,2}-\omega_c)
\tilde{\mathcal{E}}_1(r\omega_{rep,1}-\omega_c)
V_{g_2 e_1} V_{e_1 g_1}V_{g_1 e_1} V_{e_1 g_2}
\nonumber \\&&\times
\frac{\delta((r-n)\omega_{rep,1}-(p-m)\omega_{rep,2}-\omega_s)}
%\delta(n+r-m-p)}
{(r\omega_{rep,1}-\omega_{e_1g_1}+i\Gamma_{e_1g_1})
((p-m)\omega_{rep,2}-\omega_{g_2g_1}-i\Gamma_{g_2g_1})
(p\omega_{rep,2}-\omega_{e_1g_1}-i\Gamma_{e_1g_1})}
\Big].
\nonumber\\&&
\end{eqnarray}
\end{widetext}
The $\tilde{\mathcal{E}}_1^2\tilde{\mathcal{E}}_2^2$ scaling signal is designated as $1122$.
The first term
\begin{equation}
S_{t1122}^{(3)}(\omega_s)\propto
\frac{\delta((r+n)\omega_{rep,1}-(p+m)\omega_{rep,2}-\omega_s)}
%\delta(n+r-m-p)}
{((n+r)\omega_{rep,1}-\omega_{f_1g_1}+i\Gamma_{f_1})},
\end{equation}
 gives the two-photon peaks multiplied by $\eta$.
Using $n+r=m+p$ in the
delta-function gives $(n+r)\delta\omega_{rep}=\omega_s$.
Substituting $(n+r)=\omega_s/\delta\omega_{rep}$ into the denominator  yields the
 TPA resonance at $ \omega_s= \eta \omega_{f_1g_1}$.
A similar effect occurs for the second term with the Raman resonances.
From the combination of the terms
\begin{equation}
S_{t1122}^{(3)}(\omega_s)\propto
\frac{\delta((r-n)\omega_{rep,1}-(p-m)\omega_{rep,2}-\omega_s)}
%\delta(n+r-m-p)}
{((p-m)\omega_{rep,2}-\omega_{g_2g_1}-i\Gamma_{g_2g_1})},
\end{equation}
the selection $r-n=p-m$  in the delta function gives  $(p-m)\delta\omega_{rep}=\omega_s$.
Substituting this combination into
 dominator, we find the Raman resonance at
 $ \omega_s=\eta \omega_{g_2g_1}$.

%%%%%%%%%%%%%%%%%%%%%%%%%%%%%%%%%%
\begin{figure}[t]
\begin{center}
\includegraphics[scale=0.3]{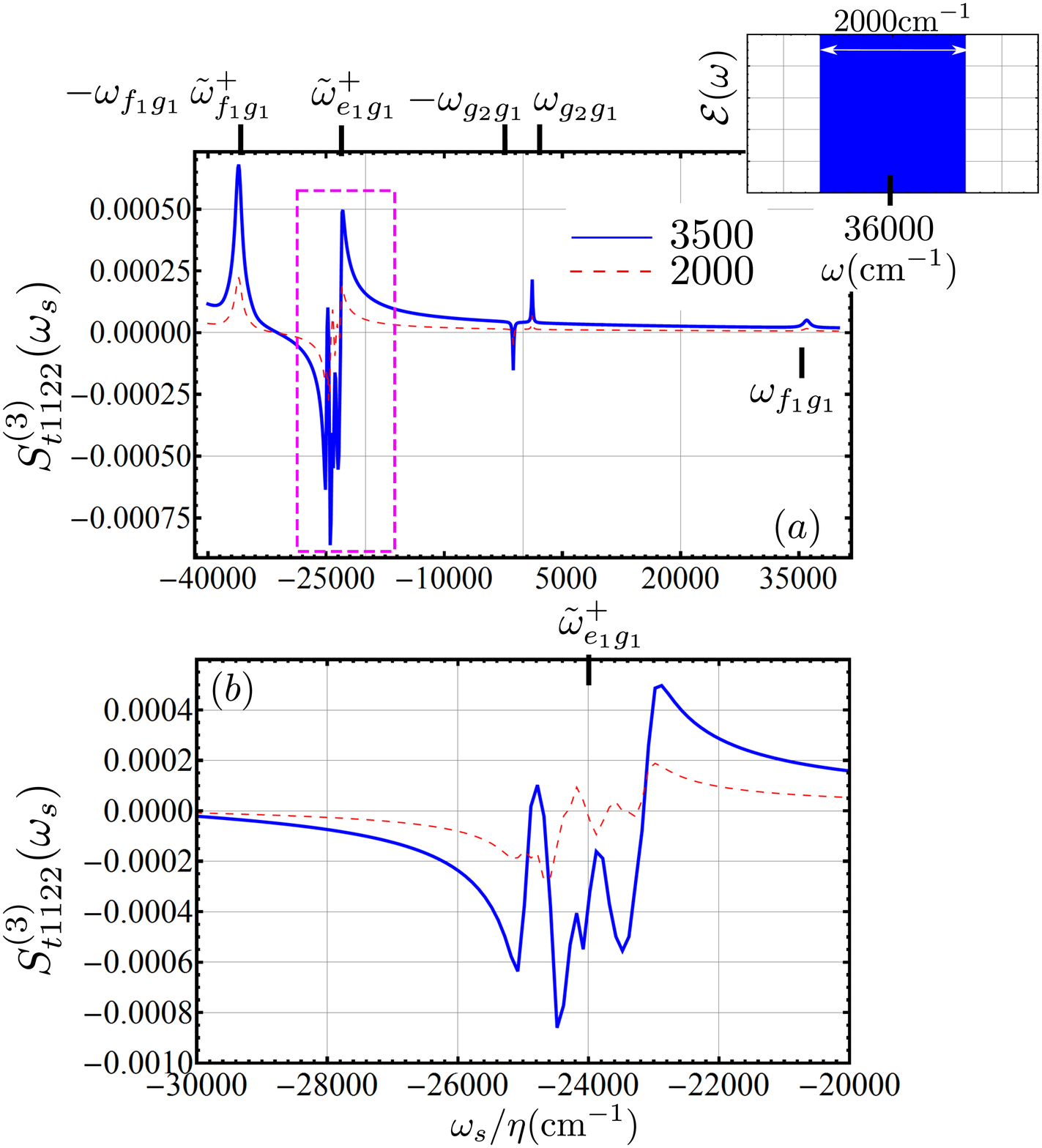}
\end{center}
\caption{(Color online)
(a) The resonant time-resolved transmission spectrum
 from Eq. \eqref{Scombo} is plotted.
Inset shows an illustration of frequency comb used,
$\omega_c=36000  \mathrm{cm}^{-1}$ and width $ 2000\mathrm{cm}^{-1}$.
%$\omega=(35000\mathrm{cm}^{-1}, 37000\mathrm{cm}^{-1})$.
(b) The spectrum in the purple region
near $\omega_s=\tilde{\omega}_{e_1 g_1}^+$
is enlarged. The expression for the peaks are given in Eqs. \eqref{TPARaman},
\eqref{peaksfig}.
%Parameters used are $\sigma=1.8*10^6\mathrm{cm}^{-1}$,
%$\omega_{rep,1}=0.080\mathrm{cm}^{-1}$,
%$\eta=10^{-6}$,
%$\Delta t=0$.
 }
 \label{fig:Plot1}
\end{figure}
%%%%%%%%%%
%%%%%%%%%%%%%%%%%%%%%%%%%%%%%%%%%%
\begin{figure}[t]
\begin{center}
\includegraphics[scale=0.3]{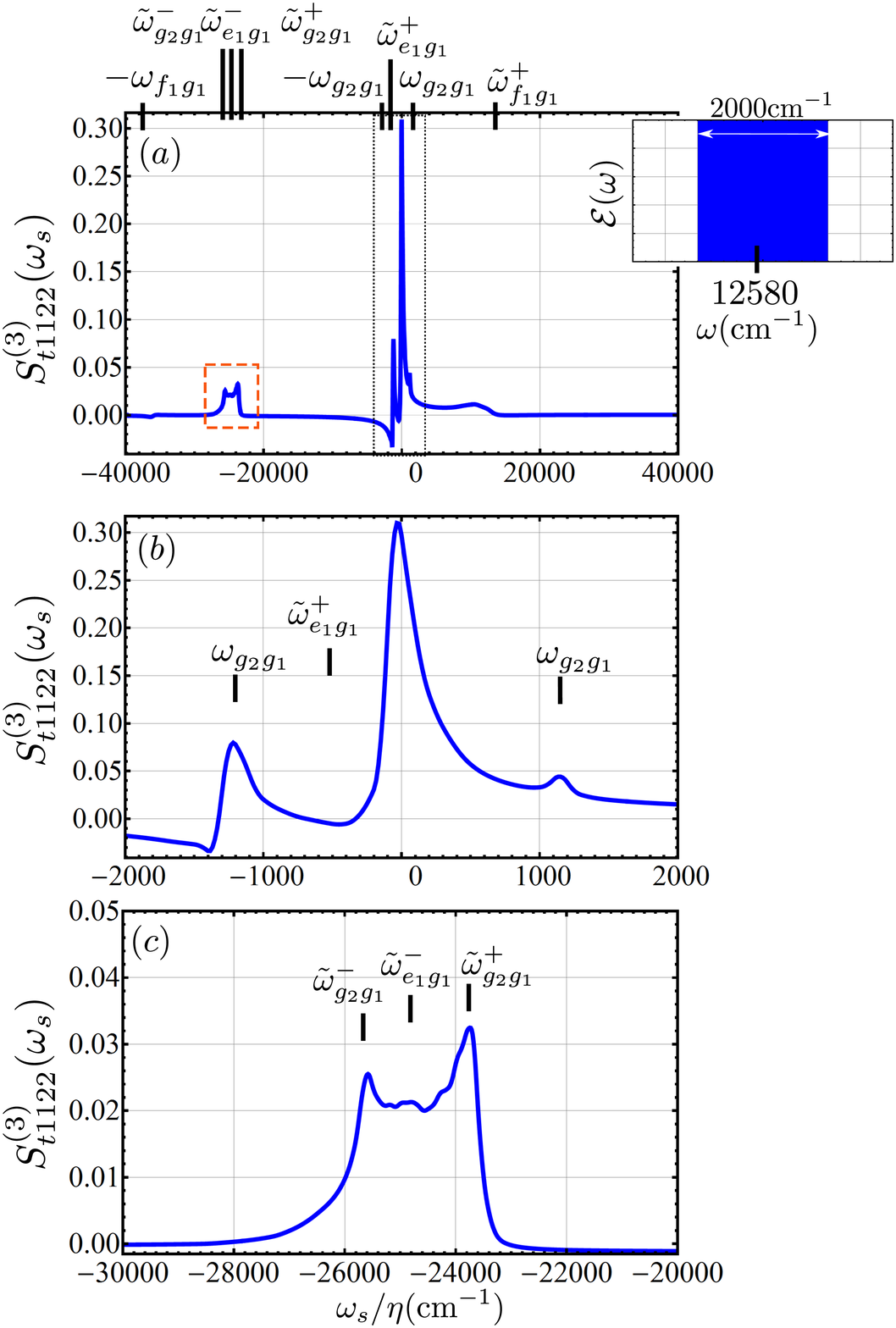}
\end{center}
\caption{(Color online)
(a) The off-resonant time-resolved transmission spectrum.
Inset shows a frequency comb centered at
$\omega_c=12580  \mathrm{cm}^{-1}$.
(b)  The spectrum for the green region about $\omega_s=0$ is enlarged.
(c)  The spectrum for the pink region near $\omega_s =\tilde{\omega}_{e_1 g_1}^-$ is enlarged.
The expression for the peaks are given in Eqs. \eqref{TPARaman},
\eqref{peaksfig}.
 }
 \label{fig:Plot12}
\end{figure}
%%%%%%%%%%

%%%%%%%%%%%%%%%%%%%%%%%%%%%%%%%%%%%%%%%%
\begin{figure*}[t]
\begin{center}
\includegraphics[width=160mm]{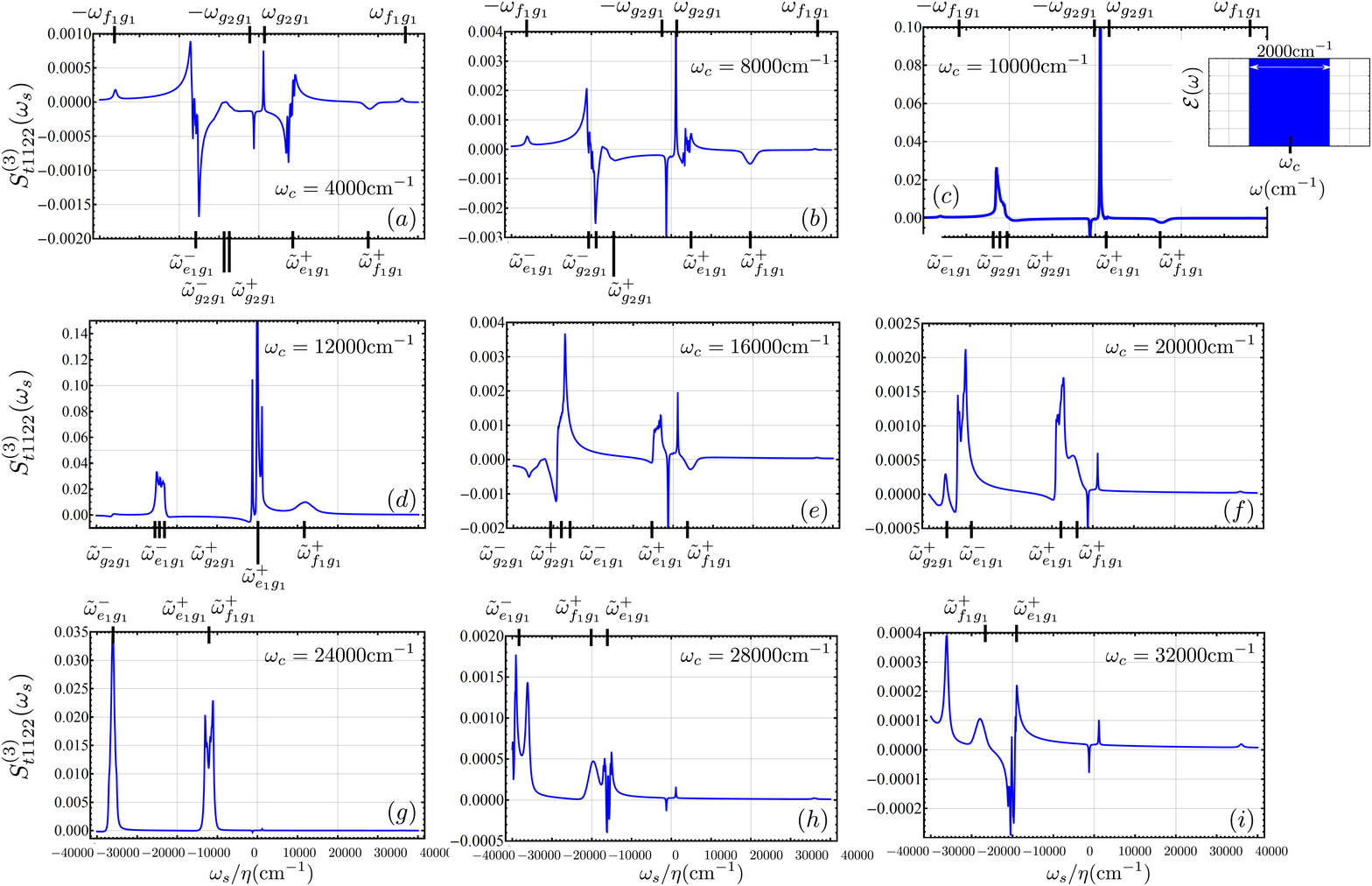}
\end{center}
\caption{(Color online)
 The time-resolved transmission spectrum $S_{t1122}^{(3)}(\omega_s)$ Eq. \eqref{Scombo} is displayed,
%for a Gaussian envelope Eq. \eqref{Gaussian} and
for
  various values of $\omega_c$.
The inset shows the frequency comb centered at
$\omega_c$ and width $2000\mathrm{cm}^{-1}$.
The expression for the peaks are given in Eqs. \eqref{TPARaman},
\eqref{peaksfig}.
}
\label{fig:manycomb}
\end{figure*}
%%%%%%%%%%%%%%%%%%%%%%%%%%%%%%%%%%%%%%%%%%%%%%%%%%%%%%%%%%%%%%%%%%%%%%%%%%%%%%%%%

Two rather large summations are required to evaluate in the  time-resolved transmission signal  Eq. \eqref{Scombo}.
%Both summations cover the same range.
We performed them using the Monte Carlo
method\cite{metropolis_monte_1949,Anderson:1986p25084}, where we randomly sample
the comb line numbers in the frequency comb. Convergence is verified my changing the sample
number and observing changes in the spectra.

%$N={22250\mathrm{cm}^{-1}/\omega_{rep}}=370833$.
The repetition frequency is selected as $\omega_{rep,1}=0.080\mathrm{cm}^{-1}$, so that the TPA peaks
can be observed within the range $\pm\omega_{rep,1}/2$. The peaks in the transmission spectra
are  multiplied by the factor $\eta=10^{-6}$, which is in the radiofrequency
range. We use a Gaussian envelope Eq. \eqref{Gaussian}
with
%$\omega_c=12580  \mathrm{cm}^{-1}$ and
$\sigma=1.8*10^6\mathrm{cm}^{-1}$.

 The resonant time-resolved transmission signal Eq. \eqref{Scombo} is displayed in Fig. \ref{fig:Plot1}(a), for $\Delta t=0$,
and $\omega_c=36000\mathrm{cm}^{-1}$.  The frequency comb range was selected as
$\omega=(35000\mathrm{cm}^{-1}, 37000\mathrm{cm}^{-1})$ and contains 25,000 comb lines.
The range was randomly sampled for  2,000 pulses (dashed-red) or 3,500 pulses (solid-blue).
 $S_{t1122}^{(3)}(\omega_s)$ is in the units $ \frac{2 \pi}{\hbar}(\frac{1}{2\pi \hbar})^3 \omega_{rep,1}^4E_1^2E_2^2$
 with the dipole moments set to one.
The inset  shows an illustration of
the frequency comb used. The spectrum shows the TPA and Raman resonances
at $\omega_s =\pm\eta \omega_{f_1g_1}$ and $\omega_s =\pm\eta \omega_{g_2g_1}$, respectively.
The  $\omega_s = \tilde{\omega}_{e_1g_1}^+$ peak
 is centered at $\omega_s= \eta( \omega_{e_1g_1}-\omega_c)=-24000\eta \mathrm{cm}^{-1}$
has a width  $\approx 2000\eta\mathrm{cm}^{-1}$.
The $\omega_s=\tilde{\omega}_{f_1g_1}^+$ peak
is located at $\omega_s=\eta(\omega_{f_1g_1}-2\omega_c)=-36000\eta \mathrm{cm}^{-1}$.
Since its position depends on both $n$ and $m$ it will have a width of  $\approx 4000\eta\mathrm{cm}^{-1}$.
This is the reason why the negative two-photon peak in Fig.  \ref{fig:Plot1}(a)
is more pronounced than the positive peak.

The shaded region  in Fig. \ref{fig:Plot1}(a) is re-plotted in Fig. \ref{fig:Plot1}(b)
 on a larger scale, which corresponds to  $\omega_s= \tilde{\omega}_{e_1g_1}^+$.
We see that the width of the peak is  $\approx 2000\eta\mathrm{cm}^{-1}$
and that it contains both absorption and emission features.
Comparing the dashed-red for 2,000 sampled pulses to the solid-blue 3,500 pulses,
we see the same features demonstrating that the data for the 2,000 sampled pulses
represents the spectrum.

The spectrum for $\omega_s>0$ in Fig. \ref{fig:Plot1}(a),  in the radiofrequency range, contains only
the Raman and TPA peaks. Compared to the frequency-dispersed transmission
spectra Fig. \ref{fig:Plot0}(c). Only the vibrational and TPA peaks are present in Fig. \ref{fig:Plot1}(b),
while the Stokes, Rayleigh, and TPA peaks are present in Fig. \ref{fig:Plot0}(c).
In Figs.  \ref{fig:Plot1}(a), \ref{fig:Plot0}(d),
the single-photon peaks are shifted by $\eta\omega_c$ or $\omega_c$ and there are Raman
resonances  not shifted by $\eta\omega_c$ or $\omega_c$.
In Fig. \ref{fig:Plot1}(a) there are TPA resonances that are not shifted by $\eta\omega_c$,
while in  Fig. \ref{fig:Plot0}(d), they are shifted by $2\omega_c$ .

The off-resonant transmission spectrum Eq. \eqref{Scombo} is shown in Fig.  \ref{fig:Plot12}(a) for
 $\omega_c=12580\mathrm{cm}^{-1}$.  The frequency comb range was selected as
$\omega=(11580\mathrm{cm}^{-1}, 13580\mathrm{cm}^{-1})$ and contains 25,000 comb lines.
The range was randomly sampled for 2,000 pulses.
The TPA peaks are very weak.  The spectrum is composed of
Raman resonances at $\omega_s =\pm \eta \omega_{g_2g_1}$, $\tilde{\omega}_{g_2g_1}^+$
$\tilde{\omega}_{g_2g_1}^-$, single-photon peaks at
$\omega_s =\tilde{\omega}_{e_1g_1}^+$, $\tilde{\omega}_{e_1g_1}^-$,
a TPA at $\omega_s =\tilde{\omega}_{f_1g_1}^+$, $-\eta\omega_{f_1g_1}$ and a  peak at $\omega_s=0$.
The peak at $\omega_s =\tilde{\omega}_{f_1g_1}^+=10840\eta\mathrm{cm}^{-1}$
has a width of  $\approx 4000 \eta\mathrm{cm}^{-1}$.

The region in the pink shaded area near $-24000 \eta\mathrm{cm}^{-1}$
  is re-plotted in  Fig.  \ref{fig:Plot12}(c)
on an expanded scale and shows the combination of the peaks
 $\omega_s=\tilde{\omega}_{e_1g_1}^-$, $\tilde{\omega}_{g_2g_1}^+$,
 and  $\tilde{\omega}_{g_2g_1}^-$,
centered at $\omega_s=-24580, -23960, -26360 \eta \mathrm{cm}^{-1}$, respectively.

The green shaded area in Fig.   \ref{fig:Plot12}(a), is replotted in Fig.  \ref{fig:Plot12}(b).
Compared to the resonant transmission spectrum Fig.  \ref{fig:Plot1}(a),
there is an additional peak at $\omega_s=0$. This peak originates from
the $\omega_s=(\omega_{e_1g_1}-n\omega_{rep,1})$, which is not multiplied
by the factor $\eta$. For $\omega_{e_1g_1}=n\omega_{rep,1}$, there is a peak
which is located within the regime $\omega_{rep,1}/2$ at zero. The
 $\omega_s=\tilde{\omega}_{e_1g_1}^+$ peak is centered
 at  $\tilde{\omega}_{e_1g_1}^+=-580 \eta \mathrm{cm}^{-1}$, with width
 $\approx 2000\eta \mathrm{cm}^{-1}$.

The spectrum for $\omega_s>0$  in Fig. \ref{fig:Plot12}(b)
contains only the Raman peak. This plot can be
compared to the experimental results of Ref. \cite{ideguchi_coherent_2013}. The spectrum
shows qualitative agreement with there findings for measuring the off-resonant
time-resolved transmission spectrum, without the peak at zero, which originates from
the single-photon peak $\omega_s=(\omega_{e_1g_1}-n\omega_{rep,1})$.
Note that we selected a repetition frequency
and level scheme
Fig. \ref{fig:level_scheme}
than Refs. \cite{ideguchi_coherent_2013}.

The time-resolved transmission spectrum for various values of $\omega_c$ are
plotted in Fig. \ref{fig:manycomb}. The inset shows the frequency comb
that we use with width $2000\mathrm{cm}^{-1}$. The  two large
summations in Eq. \eqref{Scombo} are done by randomly sampling the range with 2,000
pulses. The transmission spectrum for $\omega_c=4000\mathrm{cm}^{-1}$ is shown
in Fig. \ref{fig:manycomb}(a). It contains the two-photon and Raman resonances
at $\omega_s=\pm\eta\omega_{f_1g_1}$, $\pm\eta\omega_{g_2g_1}$ and the
five peaks in Eq. \eqref{peaksfig}. The $\omega_s=\tilde{\omega}_{e_1g_1}^+$
and $\tilde{\omega}_{e_1g_1}^-$ peaks contain both absorption and emission
features. The Raman peaks at $\omega_s=\tilde{\omega}_{g_2g_1}^+$
and $\tilde{\omega}_{g_2g_1}^-$ interfere. The   $\omega_s=\tilde{\omega}_{f_1g_1}^+$
peak is an absorption peak. Increasing $\omega_c=8000\mathrm{cm}^{-1}$,  in Fig. \ref{fig:manycomb}(b),
 shifts the five peaks in Eq. \eqref{peaksfig} toward the left. In Fig. \ref{fig:manycomb}(c),
 for $\omega_c=10000\mathrm{cm}^{-1}$ the $\omega_s=\tilde{\omega}_{e_1g_1}^+$
overlaps the  $\omega_s=\pm\eta\omega_{g_2g_1}$ peak and amplifies the peak.

In the Fig. \ref{fig:manycomb}(d), for $\omega_c=12000\mathrm{cm}^{-1}$, there is peak
located at zero, corresponding to $\omega_s=(\omega_{e_1g_1}-n\omega_{rep,1})$,
which is not multiplied by the factor $\eta$. This peak only occurs for
$\omega_{e_1g_1}=n\omega_{rep,1}=\omega_c$. In addition, the peak
$\omega_s=\tilde{\omega}_{e_1g_1}^+$ is located at zero. Increasing
$\omega_c$ further, in Fig. \ref{fig:manycomb}(f), the $\omega_s=\tilde{\omega}_{g_2g_1}^-$
position becomes located beyond the detected regime and all peaks from Eq. \eqref{peaksfig}
are located in the regime $\omega_s<0$. In Fig. \ref{fig:manycomb}(g), the $\tilde{\omega}_{e_1g_1}^-$
peak overlaps the TPA at $\omega_s=-\omega_{f_1g_1}$, amplifying the TPA.
Increasing $\omega_c$ further, it is possible to shift
the location of the $\omega_s=\tilde{\omega}_{g_2g_1}^+$ and $\tilde{\omega}_{e_1g_1}^-$
beyond the detected regime, as in Fig. \ref{fig:manycomb}(i).

%%%%%%%%%%%%%%%%%%%%%%%%%%%%%%%%%%%%%%%%%%%
%
%
%
\section{Time-resolved transmission signal with scaling $\tilde{\mathcal{E}}_1^3 \tilde{\mathcal{E}}_2$}%
\label{three}
%%%%%%%%%%%%%%%%%%%%%%%%%%%%%%%%%%%%%%%%%%%

Selection of three interactions with comb 1 and one interaction with comb 2
will give the down-converted single-photon resonances, which do not depend upon the comb line number.
The Fourier transform of the interferometric signal will give the following beat frequencies
\begin{eqnarray}
\omega_s&&=(n-r+m)\omega_{rep,1}-p\omega_{rep,2},
\nonumber\\&&
=(n+r-m)\omega_{rep,1}-p\omega_{rep,2},
\nonumber\\&&
=(n-r-m)\omega_{rep,1}+p\omega_{rep,2}.
\label{repthree}
\end{eqnarray}
Similar to the methods used in Sec. \ref{two}, We will make use a delta function to select the correct combination of line numbers in the transmission signal.
The transmission spectrum Eq. \eqref{Scombothree}  is derived in Appendix B.
The peaks in the transmission signal are
 \begin{eqnarray}
&&
\omega_{e_1g_1}=\pm\eta\omega_{e_1g_1}
\nonumber\\&&
{\Omega}_{f_1g_1}^+=\eta (\omega_{f_1g_1}-m\omega_{rep,1})
\nonumber\\&&
{\Omega}_{f_1g_1}^-=-\eta (\omega_{f_1g_1}-n\omega_{rep,1})
\nonumber\\&&
{\Omega}_{e_1g_1}^+=\eta(\omega_{e_1g_1}+(n-m) \omega_{rep})
\nonumber\\&&
{\Omega}_{e_1g_1}^-=-\eta(\omega_{e_1g_1}-(n-m) \omega_{rep})
\nonumber\\&&
{\Omega}_{g_2g_1}^+=\eta (\omega_{g_2g_1}+n\omega_{rep})
\nonumber\\&&
{\Omega}_{g_2g_1}^-=-\eta (\omega_{g_2g_1}+m\omega_{rep}).
\label{peaksfigthree}
 \end{eqnarray}
The center position of the peaks can be found by substituting $n=\omega_c/\omega_{rep,1}$
and $m=\omega_c/\omega_{rep,1}$ into Eq. \eqref{peaksfigthree}. The peak
${\Omega}_{e_1g_1}^\pm$ is centered at $\pm\omega_{e_1 g_1}$. The peak positions
 ${\Omega}_{f_1g_1}^\pm$ are shifted by $-\eta\omega_c$, while, the peaks
  ${\Omega}_{g_2g_1}^\pm$ are shifted by $\eta\omega_c$. Comparing to the peaks
in Fig. \ref{fig:Plot0}, the TPA peaks and Raman peaks in Fig. \ref{fig:Plot0}(c),
$\mathcal{S}_f(\omega)$,
are shifted by $\omega_c$, while the single-photon peaks are not shifted by
$\omega_c$.

The down-conversion of the single photon peaks can be
seen from the transmission signal Eq. \eqref{Scombothree}, which is proportional to
\begin{widetext}
\begin{eqnarray}
&&S_{t1112}^{(3)}(\omega_s; \omega_{rep,1}, \omega_{rep,2},\tau_2)
\propto
-\mathcal{I}
\frac{2}{2\pi\hbar}
\omega_{rep,1}^3
\omega_{rep,2}
\Big[
\tilde{\mathcal{E}}_1^*(m\omega_{rep,1}-\omega_c)
\tilde{\mathcal{E}}_1(n\omega_{rep,1}-\omega_c)
\tilde{\mathcal{E}}_2^*(r\omega_{rep,1}-\omega_c)
%((r+n)\omega_{rep,1}-p\omega_{rep,2}-\omega_s-\omega_c)
\nonumber \\&&\times
\tilde{\mathcal{E}}_1(p\omega_{rep,2}-\omega_c)
V_{g_1 e_1}
V_{ e_1 g_2}
V_{ g_2 e_1}
V_{ e_1 g_1}
\nonumber\\&&\times
\frac{\delta((n-r-m)\omega_{rep,1}+p\omega_{rep,2}-\omega_s)}
%\delta(n+r-m-p)}
{(p\omega_{rep,2}-\omega_{e_1g_1}+i\Gamma_{e_1})
(p\omega_{rep,2}-r\omega_{rep,1}-\omega_{g_2g_1}+i\Gamma_{g_2})
((n-r)\omega_{rep,1}+p\omega_{rep,2}-\omega_{e_1g_1}-i\Gamma_{e_1})},
\nonumber\\&&
\label{three4}
\end{eqnarray}
\end{widetext}
The index $1112$ represents the signal scaled as $\tilde{\mathcal{E}}_1^3 \tilde{\mathcal{E}}_2$.
 Equation \eqref{three4} is proportional to
\begin{eqnarray}
&&S_{t1112}^{(3)}(\omega_s; \omega_{rep,1}, \omega_{rep,2},\tau_2)
\propto
\frac{\delta((n-r-m)\omega_{rep,1}+p\omega_{rep,2}-\omega_s)}
{(p\omega_{rep,2}-\omega_{e_1g_1}+i\Gamma_{e_1})}.
\end{eqnarray}
The selection $n-r-m=-p$ from the measurement of interferometric signal gives
$p=-\omega_s/\delta\omega_{rep}$. Substituting this into the
dominator we find the single-photon resonance at $\omega_s=-\eta\omega_{e_1 g_1}$.

%%%%%%%%%%%%%%%%%%%%%%%%%%%%%%%%%%
\begin{figure}[t]
\begin{center}
\includegraphics[scale=0.25]{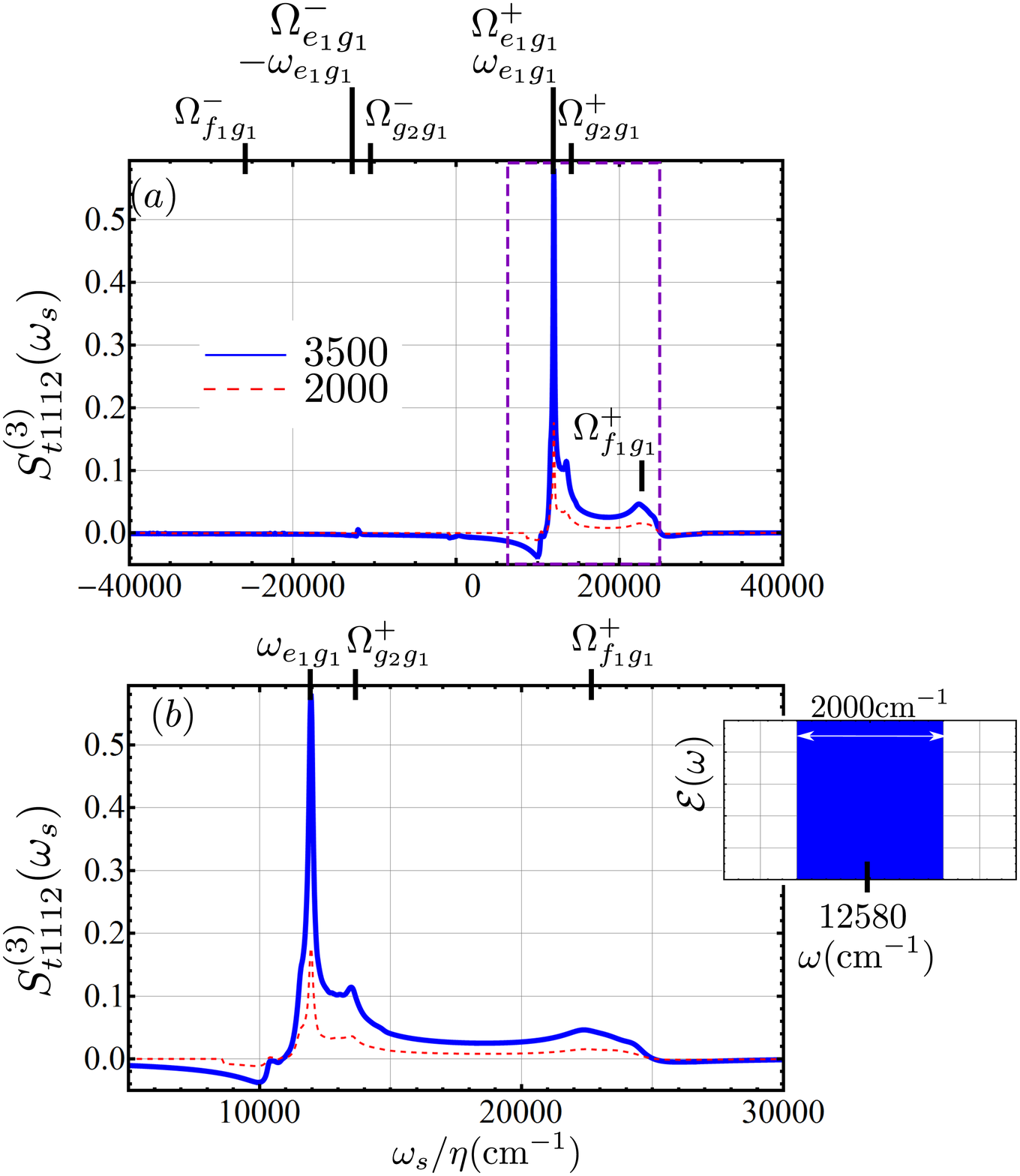}
\end{center}
\caption{(Color online)
(a) The time-resolved transmission signal for three interactions with comb 1,  Eq. \eqref{Scombothree}
is plotted. The inset shows an illustration of the frequency comb used, $\omega_c=1258 \mathrm{cm}^{-1}$.
The peaks in the spectrum are given by Eq. \eqref{peaksfigthree}. (b) The spectrum in the
purple region is re-plotted.
}
 \label{fig:Plot5a}
\end{figure}
%%%%%%%%%%

%%%%%%%%%%%%%%%%%%%%%%%%%%%%%%%%%%
\begin{figure}[t]
\begin{center}
\includegraphics[scale=0.25]{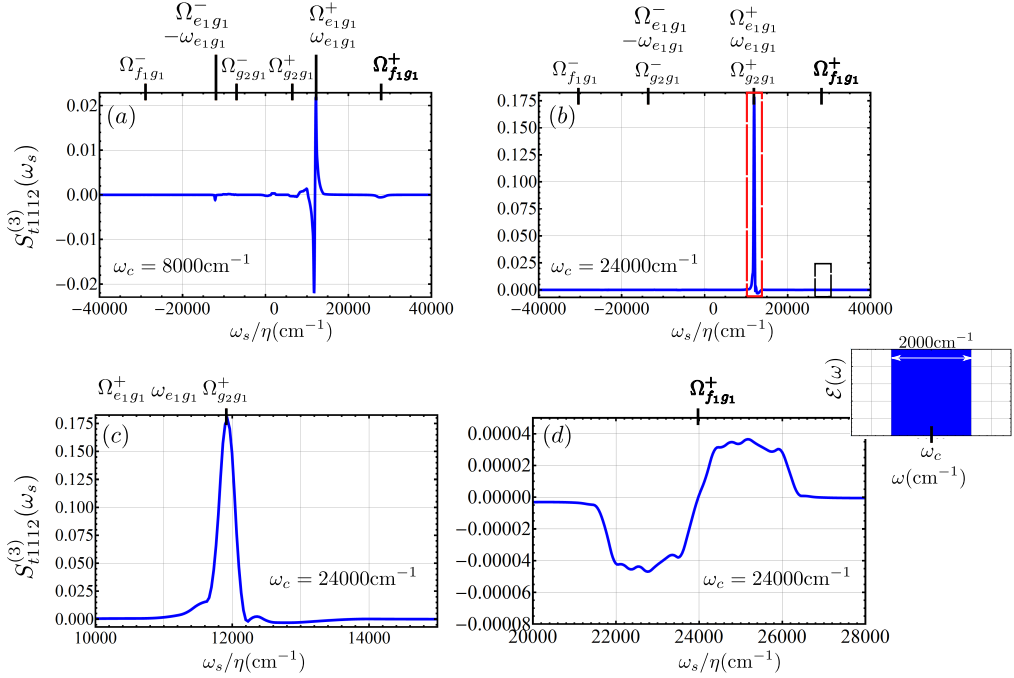}
\end{center}
\caption{(Color online)
The time-resolved transmission signal for three interactions with comb 1,  Eq. \eqref{Scombothree}
is plotted for two values of $\omega_c$: (a) $\omega_c=80000\mathrm{cm}^{-1}$,
 (b) $\omega_c=240000\mathrm{cm}^{-1}$.
 The inset shows an illustration of the frequency comb used.
The peaks in the spectrum given by Eq. \eqref{peaksfigthree}. (c) The spectrum in the
purple region of (b) is re-plotted. (d) The spectrum in the green
region of (b) is re-plotted.
}
 \label{fig:Plot7b}
\end{figure}
%%%%%%%%%%

The two large summations in the transmission signal $S_{t1112}^{(3)}(\omega_s)$ are calculated using
the Monte Carlo method, as in Sec. \ref{two}. We used the same values for the repetition frequency and  Gaussian pulse width as in Sec. \ref{two}. The off-resonant transmission signal is displayed for
$\omega_c=12580\mathrm{cm}^{-1}$ and a frequency comb width
$\omega=(11580\mathrm{cm}^{-1}, 13580\mathrm{cm}^{-1})$  in Fig. \ref{fig:Plot5a}.
The inset shows an illustration of the frequency comb used.
The range was randomly sampled for 2,000 pulses (dashed-red) and 3,500 pulses (solid-blue).
 $S_{t1112}^{(3)}(\omega_s)$ is in the units $ \frac{2 \pi}{\hbar}(\frac{1}{2\pi \hbar})^3 \omega_{rep,1}^4E_1^3E_2$ with the dipole moments set to one. The transmission spectrum
 in Fig. \ref{fig:Plot5a}(a) is dominated by the $\omega_s=\omega_{e_1 g_1}$ peak.
 The $\omega_s=\Omega^+_{e_1 g_1}$ peak has a width of $4000\eta\mathrm{cm}^{-1}$ and overlaps the $\omega_s=\omega_{e_1 g_1}$ peak.
 The $\omega_s=\Omega_{f_1 g_1}^+$ peak has a width of $2000\eta\mathrm{cm}^{-1}$.
  Comparing the 2,000 sampled to the 3,500 sampled, the features from the 2,000 pulse resemble the 3,500.
 The purple region is replotted in Fig. \ref{fig:Plot5a}(b) on a smaller regime. The
 $\omega_s=\Omega_{g_2 g_1}^+$ has a width of $2000\mathrm{cm}^{-1}$.  There is a feature
near $\omega_s=-10,000\eta\mathrm{cm}^{-1}$ that corresponds to the $\omega_s=\Omega_{e_1 g_1}^+$
peak.

The transmission signal for two values of $\omega_c$ are shown in Fig. \ref{fig:Plot7b}.
For $\omega_c<\omega_{e_1 g_1}$ in Fig. \ref{fig:Plot7b}(a), the spectrum is mostly
composed of the single-photon peak, which shows both emission and absorption features.
For $\omega_c >\omega_{e_1 g_1}$, in Fig. \ref{fig:Plot7b}(b), the single-photon peak
becomes an emission peak and all peaks dependent upon comb line number are suppressed.
There are three peaks, $\omega_s=\omega_{e_1 g_1}$, $\Omega_{g_2 g_1}^+$,  $\Omega_{e_1 g_1}^+$
that overlap at the single-photon resonance.
The purple region is replotted in Fig. \ref{fig:Plot7b}(c), showing
that the single-photon resonance has width according to the dephasing rate.
The peak at $\Omega_{f_1 g_1}^+$, the green region in Fig. \ref{fig:Plot7b}(b)
is replotted in Fig. \ref{fig:Plot7b}(d).

%%%%%%%%%%%%%%%%%%%%%%%%%%%%%%%%%%%%%%%%%%%%%%%%%%%%%%
%
%
%
%
\section{time-resolved Transmission spectra with shaped spectral phase}
%
%
%
%
%%%%%%%%%%%%%%%%%%%%%%%%%%%%%%%%%%%%%%%%%%%%%%%%%%%%%%%

The future developments in spectroscopy using the frequency comb include
shaping the individual pulses in the pulse train.
This method requires a pulse shaper to have a spectral resolution that matches the spacing
of the comb lines of the input pulse train.
This was  demonstrated recently\cite{jiang_spectral_2005,fontaine_32_2007,jiang_optical_2007,ferdous_spectral_2011,rashidinejad_generation_2013}.
Currently, this method is limited to small frequency combs, say 100 comb lines.
Generation of pulse shaping in dual comb Fourier transform spectroscopy was recently demonstrated for
triangular shaped pulses\cite{zhou_pair-by-pair_2013}. The two frequency combs contained 4 identically
shaped-pulses with slightly different repetition rates. Here, we consider pulse-shaping of a frequency comb
with 25,000 pulses using a sinusoidal spectral phase function. This was demonstrated
 in Doppler free spectroscopy\cite{barmes_spatial_2013} with a repetition
 frequency of 180 MHz (0.06$\mathrm{cm}^{-1}$).

The spectrum with two interactions with comb 1 $S_{t1122}^{(3)}(\omega)$ contains the peaks
$\omega_s=\pm \eta \omega_{g_2g_1}$ and $\omega_s=\pm \eta \omega_{f_1g_1}$, which are independent
of the comb line numbers. We are interested in controlling these resonances by means of
 employee an oscillating phase onto the pulse envelope
\begin{eqnarray}
&&\tilde{\mathcal{E}}_1(\omega)=\mathcal{E}_1(\omega) e^{i\phi(\omega)},
\hspace{0.15in}
\tilde{\mathcal{E}}_2(\omega)=\mathcal{E}_2(\omega) e^{i\phi(\omega)},
\label{PhaseProfile}
\end{eqnarray}
where $ \mathcal{E}_1(\omega) $ and  $ \mathcal{E}_2(\omega) $ represent the real part, which is
a Gaussian, Eq. \eqref{Gaussian}. The sinusoidal spectral phase reads
\begin{equation}
\phi(\omega)=\alpha \sin( \beta \omega+ \Phi),
\label{sine}
\end{equation}
where $\alpha$ is the modulation depth, $\beta$ is inverse modulation frequency and $\Phi$ is the modulation phase. A cosine spectral phase occurs when $\Phi=\pi/2$.
Adding an oscillating phase alters the temporal profile, breaking each pulse into a train of sub-pulses.

\begin{figure}[t]
\begin{center}
\includegraphics[scale=0.25]{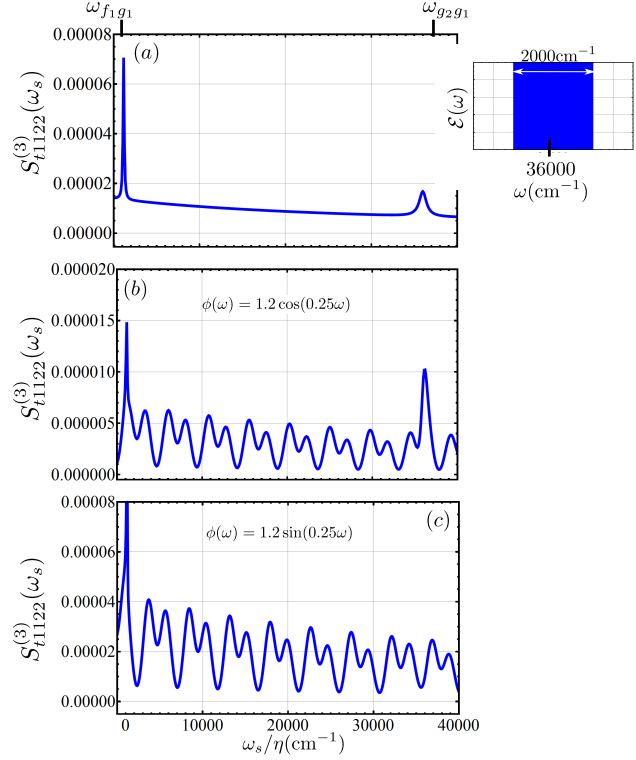}
\end{center}
\caption{(Color online)
(a) The resonant time-resolved transmission signal  Eq. \eqref{Scombo}  without a sinusoidal phase.
The resonant transmission signal   Eq. \eqref{Scombo} using  Eq. \eqref{PhaseProfile},
with an oscillating phase Eq. \eqref{sine},   for $\beta=0.25$, $\alpha=1.2$,
and two different values of the $\Phi$; (a) $\Phi=\pi/2$; (b)  $\Phi=0$.
The inset in (a) shows the frequency comb used.
}
 \label{fig:Plot3}
\end{figure}

In Fig. \ref{fig:Plot3}(a), we show the time-resolved transmission spectrum, without
an oscillating phase, for $\omega_c=36000\mathrm{cm}^{-1}$ and width $\omega=(35000\mathrm{cm}^{-1}, 37000\mathrm{cm}^{-1})$. See the inset.
For an even spectral phase  Fig. \ref{fig:Plot3}(b), $\Phi=\pi/2$, both the Raman and TPA
peaks are present. However, for an odd spectral phase   Fig. \ref{fig:Plot3}(c), $\Phi=0$,
only the Raman  peak is present.

\begin{figure}[t]
\begin{center}
\includegraphics[scale=0.25]{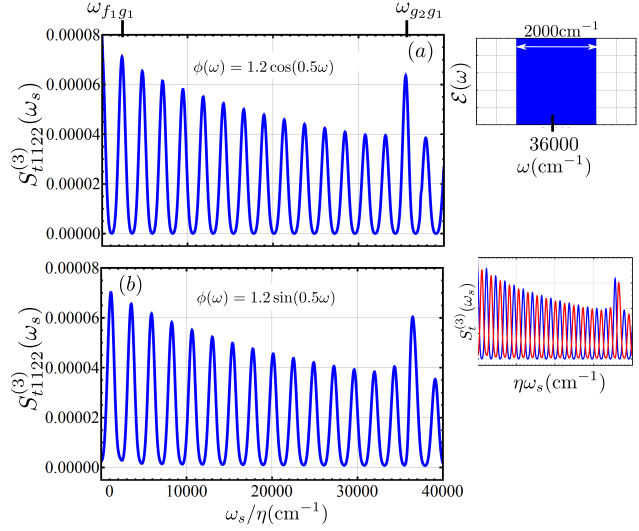}
\end{center}
\caption{(Color online)
The resonant time-resolved transmission signal   Eq. \eqref{Scombo} for $\beta=0.5$, $\alpha=1.2$,
and two different values of the $\Phi$ in  Eq. \eqref{sine}.
%We consider $\phi_1(\omega)=\phi_1(\omega)=\phi(\omega)$ in these
%plots.
(a) $\Phi=\pi/2$;
(b) $\Phi=0$.
The inset in (a) shows the frequency comb used.
}
 \label{fig:Plot4}
\end{figure}

Suppression of the Raman peak can be done by selection of the  modulation frequency $\beta$.
 In Fig. \ref{fig:Plot4}(a), for an even phase function,
the Raman peak in the spectrum is minimized while the TPA peak is enhanced.
The minimization of the Raman peak is not do to a minimum in the
oscillating spectral phase function. This is verified in Fig.  \ref{fig:Plot4}(b), where
we plot the transmission spectrum with an odd phase function $\Phi=0$. The inset, which is a plot of
(c) and (d), demonstrates that the cosine and sine spectral phase functions are out of phase.
%Note that there are oscillations in the spectrum with an oscillating phase function.

%%%%%%%%%%%%%%%%%%%%%%%%%%%%%%%%%%
%%%%%%%%%%

%%%%%%%%%%%%%%%%%%%%%%%%%%%%%%%%%%
%%%%%%%%%%

%%%%%%%%%%%%%%%%%%%%%%%%%%%%%%%%%%%%%%%%%%%%%%%%%%%%%%
%
%
%
%
\section{Summary}
%
%
%
%
%%%%%%%%%%%%%%%%%%%%%%%%%%%%%%%%%%%%%%%%%%%%%%%%%%%%%%%

We have shown that dual comb spectroscopy can be described as the time-resolved transmission signal of
 single shaped-pulse. The selection of the combination of the comb line numbers in the frequency comb
leads to  Raman, TPA and single-photon resonances in the radiofrequency regime.

For a single broadband pulse, the single-photon peaks  were shifted
by $\omega_c$. The TPA were shifted by $2\omega_c$ and the
Raman peaks are not shifted.
For the dual comb, there are several peaks in the spectrum. The time-resolved transmission signal proportional
to $\tilde{\mathcal{E}}_1^2\tilde{\mathcal{E}}_2^2$ gives single-photon peaks shifted by $\eta \omega_c$.
The TPA and Raman resonances have several peaks in the spectrum. First, peaks
 that are not shifted by $\eta \omega_c$ and have a width equal to the dephasing rate. Second,
 peaks that are shifted by   $2\eta\omega_c$ with width proportional to the width the frequency comb.
It is the selection of the
comb lines which allows some of the TPA and Raman resonances to
not be shifted by $\eta\omega_c$.

The $\tilde{\mathcal{E}}_1^3\tilde{\mathcal{E}}_2$ time-resolved transmission signal  gives TPA and Raman resonances
shifted by $\eta \omega_c$. There are two types of single-photon resonances: peaks that have a width
dependent upon the width of the frequency comb and peaks with line-widths according to the dephasing
rate.

For a  frequency comb, with several hundred thousand comb lines, the time-resolved transmission spectra will
be composed of the TPA and Raman  or single-photon resonances, which are not shifted by $\eta \omega_c$.
For a small frequency comb, with one or two comb lines, the spectra will be composed mostly of the peaks
which are shifted by $\eta\omega_c$.

\acknowledgements

We gratefully acknowledge the support of the Chemical Sciences,
Geosciences and Biosciences Division, Office of Basic Energy Sciences,
Office of Science, U.S. Department of Energy.  We also wish to
thank the National Science Foundation (Grant No. CHE-1058791)
for their support.

\appendix

%%%%%%%%%%%%%%%%%%%%%%%%%%%%%%%%%%%%%%5
%
%
%
%
%
\section{Time-resolved transmission signal--$\tilde{\mathcal{E}}_1^2 \tilde{\mathcal{E}}_2^2$}
%
%
%
%
%%%%%%%%%%%%%%%%%%%%%%%%%%%%%%%%%%%%%%%%%
Using Eq. \eqref{freq38} and the corresponding delta functions in Eq. \eqref{combo},
the transmission signal Eq. \eqref{freq00} can be cast into the following form
\begin{widetext}
\begin{eqnarray}
&&S_{t1122}^{(3)}(\omega_s ; \omega_{rep,1}, \omega_{rep,2},\Delta t)=
-\mathcal{I}
\frac{2}{\hbar}
\omega_{rep,1}^2
\omega_{rep,2}^2
\int_{-\infty}^{\infty} d\omega_1
\int_{-\infty}^{\infty} d\omega_2
\int_{-\infty}^{\infty} d\omega_3
\nonumber\\&&\times
\Big[
\tilde{\mathcal{E}}_1^*(\omega_1+\omega_2-\omega_3-\omega_s;n\omega_{rep,1})
\tilde{\mathcal{E}}_2(\omega_2;m\omega_{rep,2})
\tilde{\mathcal{E}}_2^*(\omega_3;p\omega_{rep,2})
\tilde{\mathcal{E}}_1(\omega_1;r\omega_{rep,1})
\delta(n-r-m+p)
\nonumber \\&&\times
e^{i(\omega_2-\omega_3)\Delta t}
+
\tilde{\mathcal{E}}_2^*(\omega_1+\omega_2-\omega_3-\omega_s;m\omega_{rep,2})
\tilde{\mathcal{E}}_1(\omega_2;n\omega_{rep,1})
\tilde{\mathcal{E}}_1^*(\omega_3;r\omega_{rep,1})
\tilde{\mathcal{E}}_2(\omega_1;p\omega_{rep,2})
\nonumber\\&&\times
\delta(-n+r+m-p)
e^{-i(\omega_2-\omega_3-\omega_s) \Delta t}
\Big ]
\big [
\chi^{(3)}(-\omega_1-\omega_2+\omega_3;\omega_1, \omega_2,\omega_3)
\nonumber\\&&+
\chi^{(3)}(-\omega_1-\omega_2+\omega_3; \omega_2,\omega_1,\omega_3)
\big]
+
\Big[
\tilde{\mathcal{E}}_1^*(\omega_1+\omega_2-\omega_3-\omega_s;n\omega_{rep,1})
\tilde{\mathcal{E}}_2(\omega_2;m\omega_{rep,2})
\tilde{\mathcal{E}}_1^*(\omega_3;r\omega_{rep,1})
\nonumber\\&&\times
\tilde{\mathcal{E}}_2(\omega_1;p\omega_{rep,2})
\delta(n+r-m-p)
e^{i(\omega_2 +\omega_1 )\Delta t}
+
\tilde{\mathcal{E}}_2^*(\omega_1+\omega_2-\omega_3-\omega_s;m\omega_{rep,2})
\tilde{\mathcal{E}}_1(\omega_2;n\omega_{rep,1})
\nonumber\\&&\times
\tilde{\mathcal{E}}_2^*(\omega_3;p\omega_{rep,2})
\tilde{\mathcal{E}}_1(\omega_1;r\omega_{rep,1})
\delta(-n-r+m+p)
e^{-i(\omega_1+\omega_2-\omega_s)\Delta t}
\Big]
\nonumber\\&&\times
\chi^{(3)}(-\omega_1-\omega_2+\omega_3;\omega_1, \omega_2,\omega_3)
\Big\}.
\label{freq1}
\end{eqnarray}
We used the fact that the  signal is invariant to the exchange of
$\omega_1$ and $\omega_2$ in the expressions
for the fields $\tilde{\mathcal{E}}_2(\omega_2)\tilde{\mathcal{E}}_1(\omega_1)$. The integrations over $\omega_1$, $\omega_2$ and $\omega_3$ in Eq. \eqref{freq1} can be done with the help of the delta function in the fields Eq. \eqref{freq38}, giving
\begin{eqnarray}
&&S_{t1122}^{(3)}(\omega_s ; \omega_{rep,1}, \omega_{rep,2},\Delta t)=
-\mathcal{I}
\frac{2}{\hbar}
\Big[
\mathcal{E}^*(n\omega_{rep,1}-\omega_c)
\tilde{\mathcal{E}}(m\omega_{rep,2}-\omega_c)
\tilde{\mathcal{E}}^*(p\omega_{rep,2}-\omega_c)
\tilde{\mathcal{E}}(r\omega_{rep,1}-\omega_c)
\nonumber\\&&\times
\delta(n-r-m+p)
\delta((m-p)\omega_{rep,2}+(r-n)\omega_{rep,1}-\omega_s)
e^{-i(p-m)\omega_{rep,2}\Delta t}
%\omega_3=p\omega_{rep,2}
%\omega_2=m\omega_{rep,2}
%\omega_1=r\omega_{rep,1}
\nonumber\\&&\times
\big [
\chi^{(3)}(- r\omega_{rep,1}-(m-p)\omega_{rep,2};
r\omega_{rep,1},m\omega_{rep,2},p\omega_{rep,2})
\nonumber\\&&+
\chi^{(3)}(- r\omega_{rep,1}-(m-p)\omega_{rep,2};
m\omega_{rep,2},r\omega_{rep,1},p\omega_{rep,2})
\big]
+
\mathcal{E}_2^*(m\omega_{rep,2}-\omega_c)
\nonumber\\&&\times
\tilde{\mathcal{E}}_1(n\omega_{rep,1}-\omega_c)
\tilde{\mathcal{E}}_1^*(r\omega_{rep,1}-\omega_c)
\tilde{\mathcal{E}}_2(p\omega_{rep,2}-\omega_c)
\delta(-n+r+m-p)
\nonumber\\&&\times
\delta((n-r)\omega_{rep,1}-(m-p)\omega_{rep,2}-\omega_s)
e^{-i((n-r)\omega_{rep,1}-\omega_s) \Delta t}
%\omega_3=r\omega_{rep,1}
%\omega_2=n\omega_{rep,1}
%\omega_1=p\omega_{rep,2}
\Big ]
\nonumber\\&&\times
\big [
\chi^{(3)}(-p\omega_{rep,2}-(n-r)\omega_{rep,1};
p\omega_{rep,2},n\omega_{rep,1},r\omega_{rep,1})
\nonumber\\&&+
\chi^{(3)}(-p\omega_{rep,2}-(n-r)\omega_{rep,1};
n\omega_{rep,1},p\omega_{rep,2},r\omega_{rep,1})
\big]
+
\mathcal{E}_1^*(n\omega_{rep,1}-\omega_c)
\tilde{\mathcal{E}}_2(m\omega_{rep,2}-\omega_c)
\nonumber\\&&\times
\tilde{\mathcal{E}}_1^*(r\omega_{rep,1}-\omega_c)
\tilde{\mathcal{E}}_2(p\omega_{rep,2}-\omega_c)
\delta(n+r-m-p)
\delta((m+p)p\omega_{rep,2}-(r+n)\omega_{rep,1}-\omega_s)
\nonumber\\&&\times
e^{i(m+p)\omega_{rep,2}\Delta t}
\chi^{(3)}(-(p+m)\omega_{rep,2}+r\omega_{rep,1};
p\omega_{rep,2}, m\omega_{rep,2},r\omega_{rep,1})
%\omega_3=r\omega_{rep,1}
%\omega_2=m\omega_{rep,2}
%\omega_1=;p\omega_{rep,2}
\nonumber\\&&+
\mathcal{E}_2^*(m\omega_{rep,2}-\omega_c)
\tilde{\mathcal{E}}_1(n\omega_{rep,1}-\omega_c)
\tilde{\mathcal{E}}_2^*(p\omega_{rep,2}-\omega_c)
\tilde{\mathcal{E}}_1(r\omega_{rep,1}-\omega_c)
\delta(-n-r+m+p)
\nonumber\\&&\times
\delta((n+r)\omega_{rep,1}-(m+p)\omega_{rep,2}-\omega_s)
e^{-i((n+r)\omega_{rep,1}-\omega_s)\Delta t}
%\omega_3=p\omega_{rep,2}
%\omega_2=n\omega_{rep,1}
%\omega_1=r\omega_{rep,1}
\nonumber\\&&\times
\chi^{(3)}(-(r+n)\omega_{rep,1}+p\omega_{rep,2};
r\omega_{rep,1}, n\omega_{rep,1},p\omega_{rep,2})
.
\label{freq13}
\end{eqnarray}
%Equation \eqref{freq13} contains both Raman resonances and TPA peaks.
%We can separate the two photon and Raman resonance contributions.
The last  two delta-functions can be used to eliminate two summations, giving
\begin{eqnarray}
&&S_{t1122}^{(3)}(\omega_s ;\Delta t, \delta \omega_{rep}, \omega_{rep,1})=
\sum_{n,m}^N
%3
S_{t1122}^{(3)}(\omega_s ;\Delta t, \delta \omega_{rep}, \omega_{rep,1},
n,m,
\frac{m\delta\omega_{rep}-\omega_s}
{\omega_{rep,1}}
,
\frac{n\delta\omega_{rep}-\omega_s}
{\omega_{rep,1}}
).
\nonumber\\&&
\label{Scombo}
\end{eqnarray}
where $S_{t1122}^{(3)}(\omega_s)$ is given as
\begin{eqnarray}
&&S_{t1122}^{(3)}(\omega_s ; \omega_{rep,1}, \omega_{rep,2},\Delta t,n,m,p,r)=
-\mathcal{I}
\frac{2}{\hbar}
\Big[
\mathcal{E}^*(n\omega_{rep,1}-\omega_c)
\tilde{\mathcal{E}}(m\omega_{rep,2}-\omega_c)
\tilde{\mathcal{E}}^*(p\omega_{rep,2}-\omega_c)
\nonumber\\&&\times
\tilde{\mathcal{E}}(r\omega_{rep,1}-\omega_c)
e^{-i(p-m)\omega_{rep,2}\Delta t}
\big [
\chi^{(3)}(-r\omega_{rep,1}-(m-p)\omega_{rep,2};
r\omega_{rep,1},m\omega_{rep,2},p\omega_{rep,2})
\nonumber\\&&+
\chi^{(3)}(-r\omega_{rep,1}-(m-p)\omega_{rep,2};
m\omega_{rep,2},r\omega_{rep,1},p\omega_{rep,2})
\big]
\nonumber\\&&+
\mathcal{E}^*(-m\omega_{rep,2}-\omega_c)
\tilde{\mathcal{E}}(-n\omega_{rep,1}-\omega_c)
\tilde{\mathcal{E}}^*(-r\omega_{rep,1}-\omega_c)
\tilde{\mathcal{E}}(-p\omega_{rep,2}-\omega_c)
\nonumber\\&&\times
e^{-i((-n+r)\omega_{rep,1}-\omega_s) \Delta t}
\big [
\chi^{(3)}(p\omega_{rep,2}+(n-r)\omega_{rep,1};
-p\omega_{rep,2},-n\omega_{rep,1},-r\omega_{rep,1})
\nonumber\\&&+
\chi^{(3)}(p\omega_{rep,2}+(n-r)\omega_{rep,1};
-n\omega_{rep,1},-p\omega_{rep,2},-r\omega_{rep,1})
\big]
\nonumber\\&&+
\mathcal{E}^*(n\omega_{rep,1}-\omega_c)
\tilde{\mathcal{E}}(m\omega_{rep,2}-\omega_c)
\tilde{\mathcal{E}}^*(-r\omega_{rep,1}-\omega_c)
\tilde{\mathcal{E}}(-p\omega_{rep,2}-\omega_c)
e^{i(m-p)\omega_{rep,2}\Delta t}
\nonumber\\&&\times
\chi^{(3)}((p-m)\omega_{rep,2}+r\omega_{rep,1};
-p\omega_{rep,2}, m\omega_{rep,2},r\omega_{rep,1})
+
\mathcal{E}^*(-m\omega_{rep,2}-\omega_c)
\tilde{\mathcal{E}}(-n\omega_{rep,1}-\omega_c)
\nonumber\\&&\times
\tilde{\mathcal{E}}^*(p\omega_{rep,2}-\omega_c)
\tilde{\mathcal{E}}(r\omega_{rep,1}-\omega_c)
e^{-i(-(n-r)\omega_{rep,1}-\omega_s)\Delta t}
\nonumber\\&&\times
\chi^{(3)}((-r+ n)\omega_{rep,1}+p\omega_{rep,2};
r\omega_{rep,1},- n\omega_{rep,1},p\omega_{rep,2}).
\label{S2nophase}
\end{eqnarray}

%%%%%%%%%%%%%%%%%%%%%%%%%%%%%%%%%%%%%%5
%
%
%
%
%
\section{Time-resolved transmission signal--$\tilde{\mathcal{E}}_1^3 \tilde{\mathcal{E}}_2$}
%
%
%
%
%%%%%%%%%%%%%%%%%%%%%%%%%%%%%%%%%%%%%%%%%
Inserting  Eqs. \eqref{freq38},\eqref{combo} into the transmission signal Eq. \eqref{freq00},  yields
\begin{eqnarray}
&&S_{t1112}^{(3)}(\omega_s ; \omega_{rep,1}, \omega_{rep,2},\Delta t)=
-\mathcal{I}
\frac{2}{\hbar}
\omega_{rep,1}^3
\omega_{rep,2}
\int_{-\infty}^{\infty} d\omega_1
\int_{-\infty}^{\infty} d\omega_2
\int_{-\infty}^{\infty} d\omega_3
\nonumber\\&&\times
\Big[
\tilde{\mathcal{E}}_1^*(\omega_1+\omega_2-\omega_3-\omega_s;m\omega_{rep,1})
\tilde{\mathcal{E}}_2(\omega_2;p\omega_{rep,2})
\tilde{\mathcal{E}}_1^*(\omega_3;r\omega_{rep,1})
\tilde{\mathcal{E}}_1(\omega_1;n\omega_{rep,1})
\nonumber\\&&\times
\delta(-n+r+m-p)
e^{-i(\omega_2-\omega_3-\omega_s) \Delta t}
+
\tilde{\mathcal{E}}_1^*(\omega_1+\omega_2-\omega_3-\omega_s;n\omega_{rep,1})
\tilde{\mathcal{E}}_1(\omega_2;m\omega_{rep,1})
\nonumber\\&&\times
\tilde{\mathcal{E}}_2^*(\omega_3;p\omega_{rep,2})
\tilde{\mathcal{E}}_1(\omega_1;r\omega_{rep,1})
\delta(n-r-m+p)
e^{i(\omega_2 +\omega_1 )\Delta t}
+
\tilde{\mathcal{E}}_1^*(\omega_1+\omega_2-\omega_3-\omega_s;m\omega_{rep,1})
\nonumber\\&&\times
\tilde{\mathcal{E}}_1(\omega_2;n\omega_{rep,1})
\tilde{\mathcal{E}}_1^*(\omega_3;r\omega_{rep,1})
\tilde{\mathcal{E}}_2(\omega_1;p\omega_{rep,2})
\delta(-n+r+m-p)
e^{-i(\omega_1+\omega_2-\omega_s)\Delta t}
\Big]
\nonumber\\&&
\chi^{(3)}(-\omega_1- \omega_2+\omega_3;\omega_1, \omega_2,\omega_3)
\Big\}.
\label{freq1three}
\end{eqnarray}
We used the fact that the  signal is invariant to the exchange of
$\omega_1$ and $\omega_2$ in the expressions
for the fields $\tilde{\mathcal{E}}_2(\omega_2)\tilde{\mathcal{E}}_1(\omega_1)$.
Using the delta functions in the expressions for the fields Eq. \eqref{freq38}, the integrations
over $\omega_1$, $\omega_2$ and $\omega_3$ in Eq. \eqref{freq1three} are done, giving
\begin{eqnarray}
&&S_{t1112}^{(3)}(\omega_s ; \omega_{rep,1}, \omega_{rep,2},\Delta t,n,m,p,r)=
-\mathcal{I}
\frac{2}{\hbar}
\omega_{rep,1}^3
\omega_{rep,2}
\nonumber\\&&\times
\Big[
\tilde{\mathcal{E}}_1^*(m\omega_{rep,1}-\omega_c)
\tilde{\mathcal{E}}_2(p\omega_{rep,2}-\omega_c)
\tilde{\mathcal{E}}_1^*(r\omega_{rep,1}-\omega_c)
\tilde{\mathcal{E}}_1(n\omega_{rep,1}-\omega_c)
\delta(-n+r+m-p)
\nonumber\\&&\times
\delta(-p\omega_{rep,1}+p\omega_{rep,2}-\omega_s)
e^{-i(p\omega_{rep,2}-r\omega_{rep,1}-\omega_s) \Delta t}
\nonumber\\&&\times
\chi^{(3)}((r-n)\omega_{rep,1}-p\omega_{rep,2};
n\omega_{rep,1},p\omega_{rep,2},r\omega_{rep,1})
+
\tilde{\mathcal{E}}_1^*(n\omega_{rep,1}-\omega_c)
\nonumber\\&&\times
\tilde{\mathcal{E}}_1(m\omega_{rep,1}-\omega_c)
\tilde{\mathcal{E}}_2^*(p\omega_{rep,2}-\omega_c)
\tilde{\mathcal{E}}_1(r\omega_{rep,1}-\omega_c)
\delta(n-r-m+p)
\nonumber\\&&\times
\delta(p\omega_{rep,1}-p\omega_{rep,2}-\omega_s)
e^{i(m\omega_{rep,1} +r\omega_{rep,1} )\Delta t}
\chi^{(3)}(-(r+ m)\omega_{rep,1}+p\omega_{rep,2};
r\omega_{rep,1}, m\omega_{rep,1},p\omega_{rep,2})
\nonumber\\&&+
\tilde{\mathcal{E}}_1^*(m\omega_{rep,1}-\omega_c)
\tilde{\mathcal{E}}_1(n\omega_{rep,1}-\omega_c)
\tilde{\mathcal{E}}_1^*(r\omega_{rep,1}-\omega_c)
\tilde{\mathcal{E}}_2(p\omega_{rep,2}-\omega_c)
\delta(-n+r+m-p)
\nonumber\\&&\times
\delta(-p\omega_{rep,1}+p\omega_{rep,2}-\omega_s)
e^{-i(p\omega_{rep,2}+n\omega_{rep,1}-\omega_s)\Delta t}
\nonumber\\&&\times
\chi^{(3)}(-p\omega_{rep,2}-(n-r)\omega_{rep,1};
p\omega_{rep,2},n\omega_{rep,1},r\omega_{rep,1})
\Big].
\label{S2nophasethree}
\end{eqnarray}
The last two delta-functions can be used to eliminate two of the summations, giving
\begin{eqnarray}
&&S_{t1112}^{(3)}(\omega_s ;\Delta t, \delta \omega_{rep}, \omega_{rep,1})=
\sum_{n,m}^N
%3
S_t^{(3)}(\omega_s ;\Delta t, \delta \omega_{rep}, \omega_{rep,1},
n,m,
\frac{-\omega_s}
{\delta\omega_{rep}}
,
\frac{(n-m)\delta\omega_{rep}-\omega_s}
{\delta\omega_{rep}}
).
\nonumber\\&&
\label{Scombothree}
\end{eqnarray}
where $S_{t1112}^{(3)}(\omega_s)$ is given as
\begin{eqnarray}
&&S_{t1112}^{(3)}(\omega_s ; \omega_{rep,1}, \omega_{rep,2},\Delta t,n,m,p,r)=
-\mathcal{I}
\frac{2}{\hbar}
\omega_{rep,1}^3
\omega_{rep,2}
\nonumber\\&&\times
\Big[
\tilde{\mathcal{E}}_1^*(m\omega_{rep,1}-\omega_c)
\tilde{\mathcal{E}}_2(p\omega_{rep,2}-\omega_c)
\tilde{\mathcal{E}}_1^*(r\omega_{rep,1}-\omega_c)
\tilde{\mathcal{E}}_1(n\omega_{rep,1}-\omega_c)
%\delta(-n+r+m-p)
%\delta(-p\omega_{rep,1}+p\omega_{rep,2}-\omega_s)
e^{-i(p\omega_{rep,2}-r\omega_{rep,1}-\omega_s) \Delta t}
\nonumber\\&&\times
\chi^{(3)}((r-n)\omega_{rep,1}-p\omega_{rep,2};
n\omega_{rep,1},p\omega_{rep,2},r\omega_{rep,1})
+
\tilde{\mathcal{E}}_1^*(n\omega_{rep,1}-\omega_c)
\tilde{\mathcal{E}}_1(m\omega_{rep,1}-\omega_c)
\nonumber\\&&\times
\tilde{\mathcal{E}}_2^*(-p\omega_{rep,2}-\omega_c)
\tilde{\mathcal{E}}_1(r\omega_{rep,1}-\omega_c)
%\delta(n-r-m-p)
%\delta(p\omega_{rep,1}-p\omega_{rep,2}-\omega_s)
e^{i(m\omega_{rep,1} +r\omega_{rep,1} )\Delta t}
\nonumber\\&&\times
\chi^{(3)}(-(r+m)-p\omega_{rep,2};
r\omega_{rep,1}, m\omega_{rep,1},-p\omega_{rep,2})
+
\tilde{\mathcal{E}}_1^*(m\omega_{rep,1}-\omega_c)
\tilde{\mathcal{E}}_1(n\omega_{rep,1}-\omega_c)
\nonumber\\&&\times
\tilde{\mathcal{E}}_1^*(r\omega_{rep,1}-\omega_c)
\tilde{\mathcal{E}}_2(p\omega_{rep,2}-\omega_c)
%\delta(-n+r+m-p)
%\delta(-p\omega_{rep,1}+p\omega_{rep,2}-\omega_s)
e^{-i(p\omega_{rep,2}+n\omega_{rep,1}-\omega_s)\Delta t}
\nonumber\\&&\times
\chi^{(3)}(-p\omega_{rep,2}-(n-r)\omega_{rep,1};
p\omega_{rep,2},n\omega_{rep,1},r\omega_{rep,1})
\Big].
\label{S2nophasethree2}
\end{eqnarray}

\end{widetext}

%\bibliographystyle{ieeetr}
%\bibliography{My_Library}

\bibliographystyle{ieeetr}

\end{document}